\documentclass[12pt]{article}
\usepackage[dvips]{geometry}
\usepackage[retainorgcmds]{IEEEtrantools}
\usepackage{texshade,graphicx}
\usepackage{amsmath,amsthm,cancel,mathrsfs}
\usepackage{caption}
\usepackage{subcaption}

\DeclareMathAlphabet{\mathpzc}{OT1}{pzc}{m}{it}

\newcommand{\cl}{\mathcal{C}\ell}

\newcommand{\actson}{\
					 \begin{xy}
						{\ar@{->}@/_{9pt}/(4.9,1.5);(5.5,-.6)} 
				          \end{xy}  
				     \ }
\newcommand{\bb}{\boldsymbol}

\newcommand{\m}{\mathbb}

\newcommand{\whector}[1]{\overset{\rightharpoonup}{#1}}
\newcommand{\half}{\tiny\frac12}

\makeatletter
\newcommand{\vast}{\bBigg@{4}}
\newcommand{\Vast}{\bBigg@{5}}

\makeatother

\let\phi\varphi
\DeclareGraphicsExtensions{.eps,.pdf,.png,.jpg}

\begin{document}
\bibliographystyle{alpha}
\numberwithin{equation}{section}

\newcounter{thm}
\newcounter{lemma}
\newcounter{remark}
\numberwithin{remark}{subsection}
\newtheorem{theorem}{Theorem}[section]
\newtheorem{proposition}[theorem]{Proposition}
\newtheorem{lemma}[theorem]{Lemma}
\newtheorem{cor}[theorem]{Corollary}
\newtheorem{conjecture}[theorem]{Conjecture}

\theoremstyle{definition}
\newtheorem{definition}[theorem]{Definition}
\newtheorem{example}[theorem]{Example}
\newtheorem{xca}[theorem]{Exercise}
\newtheorem{problem}[theorem]{Problem}
\newtheorem{remark}[theorem]{Remark}
\newtheorem{properties}{Properties}

\long\def\symbolfootnote[#1]#2{\begingroup\def\thefootnote{\fnsymbol{footnote}}\footnote[#1]{#2}\endgroup}

\title{Spin Precession of Dirac Particles in Kerr Geometry}
\author{Anusar Farooqui\footnote{Department of Mathematics and Statistics, McGill University, 805 Sherbrooke Street West, Montreal QC H3A 0B9, Canada. Email: farooqui@math.mcgill.ca }}
\date{\today}

\maketitle
\begin{abstract}
We isolate and study the transformation of the intrinsic spin of Dirac particles as they propagate along timelike geodesics in Kerr geometry. Reference frames play a crucial role in the definition and measurement of the intrinsic spin of test particles. We show how observers located in the outer geometry of Kerr black holes may exploit the symmetries of the geometry to set up reference frames using purely geometric, locally-available information. Armed with these geometrically-defined reference frames, we obtain a closed-form expression for the geometrically-induced spin precession of Dirac particles in the outer geometry of Kerr black holes. We show that the spin of Dirac particles does not precess on the equatorial place of Kerr geometry; and hence, in Schwarschild geometry. 
\end{abstract}

\newpage
\tableofcontents{}

\section{Introduction}
Our goal in this paper is to isolate and study the transformation of the intrinsic spin of Dirac particles as they propagate along timelike geodesics in Kerr geometry. The motivation comes from the problem of observers located in the outer geometry of Kerr black holes trying to communicate quantum information by exchanging polarized particles. It was shown in \cite{Farooqui14} how observers located in the outer geometry of Kerr black holes may exploit the symmetries of the geometry in order to communicate information by exchanging polarized photons. In the present paper, we consider the problem of observers in Kerr geometry trying to communicate quantum information by exchanging massive spin-$\half$ particles instead. In the massive spin-$\half$ case, one encounters two problems that are not present in the photonic case covered in \cite{Farooqui14}, which we now describe.

Suppose Alice encodes information in the spin of a massive spin-$\half$ particle and sends it to Bob. First, she needs to ensure that the particle's trajectory will intersect Bob's worldline. Spin-$\half$ particles are described by spinor fields that obey the Dirac equation. It is not immediate how to relate a given solution to the Dirac equation, a spinor field, to a timelike geodesic along which the test particle propagates. Second, whereas a photon's polarization $4$-vector is parallel propagated and remains orthogonal to its $4$-velocity, there is no such propagation law for the spin vector of a Dirac particle. Following \cite{Audretsch81}, we address these two problems by using the semiclassical ansatz for the Dirac equation, which allows us to recover a timelike geodesic along which the test particle propagates and derive a propagation law for the spin vector.

Terashima and Ueda's seminal paper \cite{Terashima04} outlined a strategy for evaluating the spin precession induced by the motion of  spin-$\half$ particles in a curved spacetime; a strategy that has since been followed by numerous authors \cite{Alsing09,Alsing12,Palmer12,Lanzagorta12,Said10,Robledo11}. The strategy outlined in \cite{Terashima04} is to calculate the Wigner rotation induced by the instantaneous local Lorentz transformation relating the particle's 4-momentum at nearby events along the particle's worldline. For a particle moving along a timelike geodesic, the strategy assumes that the precession of the intrinsic spin of the particle is  determined solely by the rotation of the observer's frame with respect to a frame that is parallel propagated along the particle's geodesic worldline.\footnote{If the particle is accelerated, there is an additional term that arises from boosting the 4-momentum along the worldline of the particle.} \cite{Robledo11,Said10} follow the strategy of \cite{Terashima04} to obtain the spin precession of spin-$\half$ particles on circular orbits confined to the equatorial plane of Kerr-Newman geometry. \cite{Lanzagorta12} also follows the strategy of \cite{Terashima04} to obtain the spin precession of spin-$\half$ particles on circular and radially-infalling geodesic orbits confined to the equatorial plane of Kerr geometry. 

We take a very different approach from \cite{Terashima04}. We work directly with the Dirac equation and, following \cite{Audretsch81}, use the decomposition of the Gordon decomposition of the Dirac current to first define the spin vector of a Dirac particle. Then we introduce the semiclassical ansatz for the Dirac equation and thereby recover the geodesic along which the Dirac particle propagates; a strategy first proposed in \cite{Pauli32}. We extend the results of \cite{Audretsch81} and show that the spin vector is parallel propagated along the aforementioned geodesic to $O(\hbar^{2})$ in the semiclassical ansatz (Theorem \ref{ppo1}). We develop a new method of constructing a reference frame on purely geometric criteria that allows us to obtain the proper time-dependent rotation of the spin vector in a coordinate independent manner. Our expression for the geometrically-induced precession of the spin vector is valid for a Dirac particle propagating along an \emph{arbitrary} timelike geodesic in the outer geometry of a Kerr black hole. We also obtain an expression for the spherical curvature of the curve traced out by the spin vector, which allows us to analyze the dynamical behaivour of the spin vector with a single invariant function. Even though our approach is quite different from \cite{Terashima04}, our qualitative result that the spin vector is parallel propagated along a Dirac particle's geodesic worldline agrees with the assumption underlying their strategy. 

The rest of this paper is organized as follows. In Section \ref{Kerr}, we recall some of the salient geometric
properties of the Kerr metric that will be used in this paper; including the equations of motion for timelike geodesics and parallel propagated frames along timelike geodesics. In Section \ref{DE}, we introduce the Dirac equation, recall the Gordon decomposition of the Dirac current and define the spin vector of a Dirac particle. Next, in Section \ref{semiclassical}, we introduce the semiclassical ansatz for the Dirac equation and thereby recover the classical trajectory of a Dirac particle. We prove a propagation law for the spin vector in Kerr geometry in Section \ref{proplaw}. Then, we construct a reference frame on purely geometric criteria and define the geometrically-induced precession of the spin vector in Section \ref{measure}. In Section \ref{precession}, we provide an explicit expression for the geometrically-induced precession of the spin vector and obtain a spherical curvature invariant for the curve traced out by the spin vector. We conclude with a discussion of the findings in Section \ref{discuss}.  

\begin{remark}[Notation]
We shall reserve lower case Latin indices, $a,b,c,\dots$, for arbitrary orthonormal frames in which the metric takes the form $\eta^{ab}:=\text{diag}(1,-1,-1,\\-1)$; bracketed Latin indices, $(a),(b),(c),\dots$, for orthonormal frames that are parallel propagated along a timelike geodesic and in which the metric takes the form $\eta^{(a)(b)}:=\text{diag}(1,-1,-1,-1)$; hatted Latin indices, $\hat i, \hat j, \hat k,\dots$, for the spacelike components in an orthonormal frame; unhatted Latin indices $i,j,k,\dots$, and Greek indices, $\alpha, \beta, \gamma, \dots$, for spacetime coordinates in which the line element takes the form $ds^{2}=g_{\alpha\beta}dx^{\alpha}dx^{\beta}$. We will, on occasion, use index-free notation as follows. Given a metric $\left(g_{ij}\right)$ and vector field $\bb X=X^{i}\frac{\partial}{\partial x^{i}}$, the 1-form dual to $\bb X$ will be denoted by $\bb X^{\flat}$, whose components are given by $(X^{\flat})_{i}:=g_{ij}X^{j}$. Similarly, given a 1-form $\bb\omega:=\omega_{i}dx^{i}$, the vector field dual to $\bb\omega$ will be denoted by $\bb\omega^{\sharp}$, whose components are given by $(\omega^{\sharp})^{i} :=\omega_{j}g^{ij}.$ We shall sometimes find it convenient to use semicolons to denote covariant derivatives, e.g., $\Psi_{;\alpha}:=\nabla_{\alpha}\Psi$, whereas commas will denote ordinary partial derivatives, e.g. $f_{,\alpha}:=\frac{\partial}{\partial x^{\alpha}}f$. We shall work throughout in natural units: $G=c=1$.
\end{remark}

\section{Kerr geometry}\label{Kerr}
\begin{remark}
We collect in this section some well-known facts about the Kerr metric. Further details can be found in \cite{Farooqui14}.
\end{remark}
In Boyer-Lindquist coordinates $(x^{i})=(t, r, \mathit\vartheta, \phi)$ with $-\infty<t<+\infty$,
$r_{+}<r<+\infty$, $0\le\vartheta\le\pi$, $0\le\phi<2\pi$, the Kerr metric
takes the form
\begin{equation}
\label{lineElement}
 ds^{2}= \frac{\Delta}{\Sigma}\left(dt-a\sin^{2}\vartheta d\phi\right)^{2} - \frac\Sigma\Delta dr^{2} -\Sigma d\vartheta^{2} -\frac{\sin^{2}\vartheta}\Sigma\left(adt-\left(r^{2}+a^{2}\right)d\phi\right)^{2},
\end{equation}
with
\begin{equation}
\label{sigmaDelta}
\Sigma(r, \vartheta) := r^{2}+a^{2}\cos^{2}\vartheta, \quad
\Delta(r) :=  r^{2}-2Mr+a^{2}.
\end{equation}
The parameters $M>0$ and $a\ge0$ correspond respectively to the mass and angular momentum per unit
mass of the black hole, as measured from infinity.  We shall only be considering the \emph{non-extreme case} $M>a\ge0$, which implies that the function $\Delta(r)$ has two distinct zeros,  
\begin{equation}
\label{ }
 r_{\pm}=M\pm\sqrt{M^{2}-a^{2}}.
\end{equation}
Moreover, we shall restrict our attention to the region $r>r_{+}$, which describes the geometry outside the event horizon of the black hole. 

The Kerr metric admits a two-parameter Abelian isometry group generated by the pair of commuting Killing vector fields
$\partial_{t}$ and $\partial_{\phi}$. The Kerr metric also admits a discrete subgroup isomorphic to
$\mathbb{Z}_{2}$ generated by the involutive isometry 
\begin{equation}
\label{inv}
(t,r,\vartheta, \phi)\mapsto(-t, r, \vartheta, -\phi).
\end{equation}
We shall denote by $L$ the differential of the isometry (\ref{inv}). The Weyl conformal curvature tensor of the Kerr solution is of Petrov type
D, meaning that it admits a pair of repeated principal null
directions, each of which is defined up to multiplication by a non-zero
scalar function. We eliminate the scaling freedom we would have otherwise had in defining a null coframe adapted to the principal null directions of the Weyl tensor as follows.  
\begin{definition}[Symmetric frame] 
Our null coframe is chosen such that 
 \begin{equation}
\label{invol}
L\bb\vartheta^{1}=-\bb\vartheta^{2}, \ L\bb\vartheta^{2}=-\bb\vartheta^{1}, \ L\bb\vartheta^{3}=-\bb\vartheta^{4}, \ L\bb\vartheta^{4}=-\bb\vartheta^{3}.
\end{equation}
We refer to this frame as the \emph{symmetric null coframe}. It is given in Boyer-Lindquist coordinates by
\begin{eqnarray}
\bb\vartheta^{1} & = & \frac1{\sqrt{2\Sigma\Delta}}\left(\Delta dt+\Sigma dr - a\sin^{2}\vartheta\Delta d\phi\right), \\
\bb\vartheta^{2} & = & \frac1{\sqrt{2\Sigma\Delta}}\left(\Delta dt-\Sigma dr - a\sin^{2}\vartheta\Delta d\phi\right), \\
\bb\vartheta^{3} & = & \frac1{\sqrt{2\Sigma}}\left(\left(r^{2}+a^{2}\right)\sin\vartheta d\phi -i\Sigma d\vartheta-a\sin\vartheta dt\right),\\
\bb\vartheta^{4} & = & \frac1{\sqrt{2\Sigma}}\left(\left(r^{2}+a^{2}\right)\sin\vartheta d\phi +i\Sigma d\vartheta-a\sin\vartheta dt\right).
\end{eqnarray}
\end{definition}
The \emph{orthonormal symmetric coframe} $(\bb\omega^{0},\bb\omega^{1},\bb\omega^{2},\bb\omega^{3})$ corresponding to the symmetric null coframe $(\bb\vartheta^{1},\bb\vartheta^{2},\bb\vartheta^{3},\bb\vartheta^{4})$ is then defined by 
\begin{equation}
\label{carter}
\bb\omega^{0}=\frac{1}{\sqrt{2}}(\bb\vartheta^{1}+\bb\vartheta^{2}), \bb\omega^{1}=\frac{1}{\sqrt{2}}(\bb\vartheta^{2}-\bb\vartheta^{1}), \bb\omega^{2}=-\frac{1}{\sqrt{2}}(\bb\vartheta^{3}+\bb\vartheta^{4}), \bb\omega^{3}=\frac{i}{\sqrt{2}}(\bb\vartheta^{3}-\bb\vartheta^{4}).
\end{equation} 
The principal null directions of the Weyl tensor, with scale factors chosen according to the requirement (\ref{invol}), are given by
\begin{equation}\label{lsymm1}
\bb\ell=\ell^{i}\frac{\partial}{\partial
  x^{i}}=\frac{1}{\sqrt{2\Sigma\Delta}}\left(\left(r^{2}+a^{2}\right)\frac{\partial}{\partial
    t}+\sqrt{\Delta}\frac{\partial}{\partial r}+a\frac{\partial}{\partial
    \varphi}\right), 
\end{equation}
and
\begin{equation}\label{nsymm1}
\bb n=n^{i}\frac{\partial}{\partial
  x^{i}}=\frac{1}{\sqrt{2\Sigma\Delta}}\left(\left(r^{2}+a^{2}\right)\frac{\partial}{\partial
    t}-\sqrt{\Delta}\frac{\partial}{\partial r}+a\frac{\partial}{\partial
    \varphi}\right).  
\end{equation}
We now define observers in terms of the principal null directions of the Weyl tensor and the involution $L$. 
\begin{definition}[Carter observers]\label{CarterObserversDef}
The vector field, 
\begin{equation}
\label{Observers}
\bb O:=\frac{1}{\sqrt{2}}(\bb\ell+\bb n)=\frac{1}{\sqrt{\Sigma\Delta}}\left(\left(r^{2}+a^{2}\right)
\frac{\partial}{\partial t}+a\frac{\partial}{\partial \varphi}\right), 
\end{equation}
where $\bb\ell$ and $\bb n$ are given by (\ref{lsymm1}) and (\ref{nsymm1}), is timelike and future-pointing, and
identifies a family of observers, that we call \emph{Carter observers}, whose 4-velocities are symmetric linear
combinations of the principal null directions of the Weyl tensor.

Carter observers exist everywhere outside the event horizon
\emph{including} the region between the event horizon and the stationary limit surface where the stationary Killing field $\bb\xi=\partial_{t}$ becomes null. Their angular velocity is $\frac{a}{r^{2}+a^{2}}$, which is exactly
the angular velocity of the event horizon; both as measured at infinity.
Therefore, this class of observers is uniquely suited to analyze the
behaviour of test particles near the horizon.

We choose the observers' frames to be dual to the symmetric coframe defined by (\ref{carter}). These frames can constructed using locally-available geometric data. Specifically, the construction of the symmetric frame only requires knowledge of the principal null directions of the Weyl tensor.  

\end{definition}


In addition to its two-parameter Abelian group of isometries and the involutive isometry, the Kerr metric posesses further symmetries that are ``hidden'' in the sense that they cannot be represented by Killing vector fields. The existence of these hidden symmetries is closely tied to the fact that all the known massless and massive wave equations are separable.  The geometric object that generates all these additional symmetries is a rank two \emph{Killing-Yano tensor}, that is, a $(0,2)$ skew-symmetric tensor $(f_{ij})$ satisfying the Killing-Yano equation, 
\begin{equation}
\label{KYE}
\nabla_{(i}f_{j)k}=0.
\end{equation}
In Boyer-Lindquist coordinates and in the symmetric orthonormal coframe, 
\begin{equation}
\label{KYT}
\bb f=-a\cos\vartheta\bb\omega^{0}\wedge\bb\omega^{1}+r\bb\omega^{2}\wedge\bb\omega^{3},
\end{equation}
is a Killing-Yano 2-form. The symmetric (0,2)-tensor $(K_{ij})$ defined by
\begin{equation}
\label{ktdef}
K_{ij}=f_{ik}f^{k}_{{\phantom k}j},
\end{equation}
satisfies the Killing equation
\begin{equation}
\label{killingeqn}
\nabla_{(i}K_{jk)}=0,
\end{equation}
and therefore gives rise to a quadratic first integral for the geodesic flow in Kerr geometry first discovered by \cite{Carter68h},
\begin{equation}
\label{fourthintegral}
\kappa=K_{ij}U^{i}U^{j},
\end{equation}
where $\left(U^{i}\right)$ is the 4-velocity. The quadratic first integral defined by equation (\ref{fourthintegral}) exists in addition to the two linear first integrals arising from the presence of the two commuting Killing vector fields $\partial_{t}$ and $\partial_{\phi}$ and therefore reduces the integration of the geodesic flow to quadratures. 
The equations of motion for timelike geodesics are given in Boyer-Lindquist coordinates by
\begin{eqnarray}
\label{geo3}
\Sigma\Delta\dot t &=& \left(\left(r^{2}+a^{2}\right)^{2}-a^{2}\Delta\sin^{2}\vartheta\right)E-2Mra\Phi,\\
\label{geo4}
\Sigma\Delta\dot\varphi &=& 2MraE +\frac{\Delta-a^{2}\sin^{2}\vartheta}{\sin^{2}\vartheta}\Phi,\\
\label{geo7}
\Sigma^{2}\dot r^{2} &=& R(r),\\
\label{geo8}
\Sigma^{2}\dot\vartheta^{2} &=& \Theta(\vartheta),
\end{eqnarray}
where
\begin{eqnarray}
\label{defR}
R(r)&:=&\m P^{2}-\Delta\left(\kappa+r^{2}\right),\\
\label{defTheta}
\Theta(\vartheta)&:=&\kappa-a\cos^{2}\vartheta - \m D^{2},
\end{eqnarray}
and 
\begin{eqnarray}
\m P(r)&:=&E\left(r^{2}+a^{2}\right)-a\Phi,\\
\m D(\vartheta) &:=&aE\sin\vartheta-\frac{\Phi}{\sin\vartheta}.
\end{eqnarray}
In equations (\ref{geo3}-\ref{geo8}), $E:=p_{t}$ is the conserved energy, $\Phi:=-p_{\varphi}$ is the conserved angular momentum, and the affine parameter is chosen such that the 4-velocity $\left(U^{i}\right)$ has unit norm, $g_{ij}U^{i}U^{j}=+1$.
The 4-velocity $\left(U^{i}\right)$ of an arbitrary timelike geodesic is given in the symmetric coframe by
\begin{equation}
\label{ }
\bb U:=\frac1{\surd\Sigma}\left(\frac{\m P}{\surd{\Delta}}, \frac{\surd R}{\surd\Delta}, \m D, \surd\Theta\right).
\end{equation}
In order to obtain an orthonormal frame $L_{(a)}$ that is parallel propagated along $\bb U$, we recall Marck's elegant construction \cite{Marck83}. 

Note first that $\bb U$ is parallel propagated along itself so that we can set 
\begin{equation}
\label{pp0}
L_{(0)}^{\phantom{(0)}a}:=U^{a}.
\end{equation}
%

The Killing-Yano 2-form $\bb f$, defined by equation (\ref{KYE}) and given in the symmetric frame by equation (\ref{KYT}), gives rise to a spacelike vector $\left(L^{a}\right)$, 
\begin{equation}
\label{kypp}
L^{a}: = f^{a}_{\phantom{a}b}U^{b},
\end{equation}
that is parallel propagated along $\bb U$ as a direct consequence of the Killing-Yano equation. We normalize $L$, defined by equation (\ref{kypp}), to obtain a unit-norm vector given by
\begin{equation}
\label{pp3}
L_{(3)}:= \frac1{\sqrt{\kappa\Sigma}}\left(\frac{a\cos\vartheta\surd R}{\surd\Delta}, \frac{a\cos\vartheta\m P}{\surd\Delta}, r\surd\Theta, -r\m D\right).
\end{equation}
In order to obtain the remaining two vectors of the parallel propagated frame, we start with a basis
\begin{eqnarray}
\widetilde L_{(1)} & = &  \frac1{\sqrt{\kappa\Sigma}}\left(\frac{\varpi r \surd R}{\surd\Delta}, \frac{\varpi r\m P}{\surd\Delta}, -\frac{a\cos\vartheta\surd\Theta}\varpi, \frac{a\cos\vartheta\m D}\varpi\right), \\
\widetilde L_{(2)} & = &  \frac1{\sqrt{\Sigma}}\left(\frac{\varpi \m P}{\surd\Delta}, \frac{\varpi \surd R}{\surd\Delta}, \frac{\m D}{\varpi}, \frac{\surd\Theta}{\varpi}\right),
\end{eqnarray}
where 
\begin{equation}
\label{varpi}
\varpi:=\frac{\kappa - a^{2}\cos^{2}\vartheta}{r^{2}+\kappa},
\end{equation}
Then we solve the ODE that governs a proper time dependent rotation angle $\Phi(\tau)$ such that the two vectors 
\begin{eqnarray}
\label{pp1}
L_{(1)} & = & \cos\Phi(\tau) \widetilde\lambda_{(1)}-\sin\Phi(\tau)\widetilde\lambda_{(2)}, \\
\label{pp2}
L_{(2)} & = & \sin\Phi(\tau) \widetilde\lambda_{(1)}+\cos\Phi(\tau)\widetilde\lambda_{(2)},
\end{eqnarray}
are parallel propagated along $\bb U$. The solution, originally obtained in \cite{Marck83}, is given by
\begin{equation}
\label{ }
\frac{d\Phi}{d\tau} = \frac{\kappa^{\half}}{\Sigma}\left[\frac{\m P}{r^{2}+\kappa} - a\sin\vartheta\frac{\m D}{\kappa-a^{2}\cos^{2}\vartheta}\right],
\end{equation}
where $\tau$ denotes proper time. The proper time dependent angle $\Phi$ may be obtained by separation of variables, 
\begin{equation}
\label{ }
\Phi(r,\vartheta):= F(r)+G(\vartheta),
\end{equation}
where
\begin{equation}
\label{ }
F(r) := \kappa^{\half}\int^{r}\frac{\m P}{r^{2}+\kappa}\frac{dr}{\surd R}, \ \ G(\vartheta) = a\kappa^{\half}\int^{\vartheta}\frac{\sin\vartheta\m D}{\kappa-a^{2}\cos^{2}\vartheta}\frac{d\vartheta}{\surd\Theta}. 
\end{equation}

\begin{theorem}[Marck]
Let $\gamma$ be a timelike geodesic parametrized by proper time $\tau$, with tangent vector $U:=\frac{d}{d\tau}\gamma$. Then, the orthonormal frame, $L_{(a)}$, defined by equations (\ref{pp0}), (\ref{pp1}), (\ref{pp2}), and (\ref{pp3}), is parallel propagated along $\gamma$. 
\end{theorem}

\section{The Dirac equation}\label{DE}
\noindent The Dirac equation is given by
\begin{equation}
\label{Deqn1}
i\hbar\gamma^{ a}\nabla_{ a}\Psi = m\Psi,
\end{equation}
where $\Psi$ is a 4-component spinor field, $m>0$ is the rest mass of the Dirac particle, $\nabla_{a}$ is the covariant derivative, $\hbar$ is Planck's constant, and $\gamma^{a}$ is a representation of the Clifford algebra $\cl_{1,3}(\m R)$, 
\begin{equation}
\label{cliff}
\gamma^{ a}\gamma^{ b}+\gamma^{b}\gamma^{a}=2\eta^{ a b}\bb1,
\end{equation}
where $\bb1$ is the unit element of the algebra. 
As a direct consequence of the Clifford algebra relations (\ref{cliff}), the commutators of the gamma matrices, 
\begin{equation}
\label{sigrep}
\sigma^{ab}:=\frac{i}2\left[\gamma^{ a}, \gamma^{ b}\right], 
\end{equation}
constitute a representation of the Lie algebra $\mathfrak{sl}(2,\m C)$ defined by the relations 
\begin{equation}
\label{Lorentzalgebra}
\left[\sigma^{ab}, \sigma^{cd}\right]=i\left(\eta^{ad}\sigma^{bc}-\eta^{ac}\sigma^{bd}+\eta^{bc}\sigma^{ad}-\eta^{bd}\sigma^{ac}\right).
\end{equation}
The generators of the Lie algebra (\ref{sigrep}), satisfying the commutation relations (\ref{Lorentzalgebra}), will play a key role in our definition of the spin vector for Dirac particles. We shall be using the standard representation of the Clifford algebra given by 
\begin{equation}
\label{gamrep}
\gamma^{0}=\left(\begin{array}{cc}I & 0 \\0 & -I\end{array}\right), \ \ \gamma^{i}=\left(\begin{array}{cc}0 & \sigma^{i} \\ -\sigma^{i} & 0\end{array}\right) \ \  (i=1,2,3),
\end{equation}
where $I$ is the $2\times2$ identity matrix. The $2\times2$ Pauli spin matrices $\sigma^{i}$ are explicitly by
\begin{equation}
\sigma^{1}:=\left(\begin{array}{cc}0 & 1 \\1 & 0\end{array}\right), \ \ \sigma^{2}:=\left(\begin{array}{cc}0 & i \\ i & 0\end{array}\right), \ \ \sigma^{3}:=\left(\begin{array}{cc}1 & 0 \\0 & -1\end{array}\right). 
\end{equation}

Given a spinor field $\Psi$, the \emph{adjoint spinor} is defined by $\bar\Psi:=\Psi^{\dagger}\gamma^{0}$, where the dagger denotes complex conjugation and transposition. If $\Psi$ satisfies the Dirac equation (\ref{Deqn1}), then $\bar\Psi$ satisfies the \emph{adjoint Dirac equation}, 
\begin{equation}
\label{Deqn2}
i\hbar\nabla_{a}\bar\Psi\gamma^{a} = -m\bar\Psi.
\end{equation}
For a spinor field $\Psi$ satisfying the Dirac equation (\ref{Deqn1}), the quantity 
\begin{equation}
\label{dcurrent}
j^{a}:=\bar\Psi\gamma^{a}\Psi,
\end{equation}
defines a vector field called the \emph{Dirac current}. The Dirac current is conserved, that is, 
\begin{equation}
\label{concurrent}
 \nabla_{a}j^{a}=0,
\end{equation}
as can be seen by multiplying equation (\ref{Deqn1}) on the left by $\bar\Psi$, multiplying equation (\ref{Deqn2}) on the right by $\Psi$, and adding the two terms together to eliminate the mass term. 

We now turn to Gordon's result \cite{Gordon28}, which demonstrates that the Dirac current decomposes into two parts which are separately conserved. 

\begin{theorem}[Gordon decompositon of the Dirac current]\label{GordonDecomp}
The Dirac current decomposes into convection and polarization 4-currents which are separately conserved. More precisely, 
\begin{equation}
\label{Gordon}
\bar\Psi\gamma^{ a}\Psi=:j^{a}=j^{a}_{polar}+j_{con}^{a},
\end{equation}
where
\begin{equation}
\label{polarcurrent}
j_{polar}^{a}:=\frac{\hbar}{2m}\left(\bar\Psi\sigma^{ab}\Psi\right)_{;b},
\end{equation}
is the polarization current satisfying $\nabla_{a}j_{polar}^{a}=0$ and 
\begin{equation}
\label{convection}
j_{con}^{a}:=\frac{\hbar}{2mi}\left(\bar\Psi^{;a}\Psi-\bar\Psi\Psi^{;a}\right),
\end{equation}
is the convection 4-current satisfying $\nabla_{a}j_{con}^{a}=0$.
\begin{proof}
The decomposition follows from noting that 
\begin{eqnarray}
\frac2{i}\left(\bar\Psi\sigma^{ab}\Psi\right)_{;b} & = & \left(\bar\Psi\gamma^{a}\gamma^{b}\Psi\right)_{;b} -\left(\bar\Psi\gamma^{b}\gamma^{a}\Psi\right)_{;b} \\
& = & \frac{2m}{i\hbar}j^{a}+\bar\Psi_{;b}\gamma^{a}\gamma^{b}\Psi -\bar\Psi\gamma^{b}\gamma^{a}\Psi_{;b}\\
& = & \frac{4m}{i\hbar}j^{a}+2\left(\bar\Psi^{;a}\Psi-\bar\Psi\Psi^{;a}\right),
\end{eqnarray}
where we have used the Dirac equation and its conjugate repeatedly, along with the symmetrization and antisymmetrization of the Dirac gamma matrices. The conservation of the convection current follows from recalling Lichnerowicz' identity \cite{Lichnerowicz64}, 
\begin{equation}
\Delta = -\nabla^{a}\nabla_{a}+\frac14R, 
\end{equation}
where $R$ is the scalar curvature and $\Delta$ is the Laplace operator whose action on the Dirac spinor and its adjoint is given by
\begin{equation}
\Delta\Psi = \gamma^{a}\gamma^{b}\nabla_{a}\nabla_{b}\Psi=\left(\frac{m}\hbar\right)^{2}\Psi, 
\end{equation}
and 
\begin{equation}
\Delta\bar\Psi=\nabla_{a}\nabla_{b}\bar\Psi\gamma^{a}\gamma^{b} = \left(\frac{m}\hbar\right)^{2}\bar\Psi, 
\end{equation}
respectively. It follows that \begin{eqnarray}
\left(\bar\Psi^{;a}\Psi-\bar\Psi\Psi^{;a}\right)_{;a} &=& \left(\nabla^{a}\nabla_{a}\bar\Psi\right)\Psi-\bar\Psi\left(\nabla^{a}\nabla_{a}\Psi\right)\\
\label{kgeqn2}
&=& \frac14R\bar\Psi\Psi-\left(\Delta\bar\Psi\right)\Psi - \frac14R\bar\Psi\Psi+\bar\Psi\left(\Delta\Psi\right),\\
&=& \left(\frac{m}\hbar\right)^{2}\left(-\bar\Psi\Psi+\bar\Psi\Psi\right)\\
&=&0.
\end{eqnarray}
Since both the Dirac current and the convection current are conserved, the polarization current, defined by equation (\ref{polarcurrent}), must be conserved as well. 
\end{proof}
\end{theorem}

For a given Dirac spinor $\Psi$, the convection current $j^{a}_{con}$ defines a congruence of timelike curves with unit tangent vector $\left(K^{a}\right)$ defined by:
\begin{equation}
K^{a}:= \frac{j^{a}_{con}}{\sqrt{\eta_{ab}j^{a}_{con}j^{b}_{con}}}.
\end{equation}

We are now in a position to define the spin vector. The following definition was originally proposed in \cite{Audretsch81}.

\begin{definition}[Spin vector]\label{spinvectordef}
We define the spin vector associated to the spinor $\Psi$ by
\begin{equation}
\label{def}
W^{a} := \frac12\varepsilon^{abcd}K_{b}\frac{\bar\Psi\sigma_{cd}\Psi}{\bar\Psi\Psi},
\end{equation}
where $\varepsilon^{abcd}$ is the anti-symmetric Levi-Civita symbol given by
\begin{equation}
\varepsilon^{abcd}:=\left\{
\begin{array}{rl}
+1 & \text{if } abcd \text{ is an even permutation of 0123},\\
-1 & \text{if } abcd \text{ is an odd permutation of 0123},\\
0 & \text{for repeated indices}.
\end{array}\right.
\end{equation}
\end{definition}
Since the $\sigma$-matrices, defined by equation (\ref{sigrep}), are the generators of the Lie algebra (\ref{Lorentzalgebra}), we may interpret $\bar\Psi\sigma^{ab}\Psi$ as spin density. Kirsch et al. have shown that the spin operator defined uniquely via the Gordon decomposition corresponds to the Foldy-Wouthuysen mean-spin operator \cite{Kirsch01}. Meanwhile, Bauke et al. have recently shown that, in the Lagrangian formulation, the Foldy-Wouthuysen mean-spin operator is the only known relativistic spin operator that commutes with the free Dirac Hamiltonian, has the eigenvalues $\pm\frac\hbar2$, and obeys the angular momentum algebra \cite{Bauke14}. Similar results were derived in \cite{Caban13}, who further show that the Foldy-Wouthuysen mean-spin operator is the only spin operator proposed so far, that has the right non-relativistic limit and does not convert positive (negative) energy states into negative (positive) energy states (their ``charge symmetry condition''). These results give us good confidence that the definition (\ref{spinvectordef}), first proposed in \cite{Audretsch81}, of the spin vector in terms of the Gordon decomposition of the Dirac current is a good one. 

We now turn to the semiclassical ansatz which will allow us to recover the classical trajectory of a Dirac particle and derive a propagation law for the spin vector defined by (\ref{def}).

\section{Semiclassical ansatz}\label{semiclassical}
\noindent The semiclassical ansatz for the Dirac equation is a formal power series expansion in Planck's constant $\hbar$, and given explicitly by
\begin{equation}
\label{ansatz}
\Psi(x)=\exp(iS(x)/\hbar)\sum_{n=0}^{\infty}\left(-i\hbar\right)^{n}a_{n}(x).
\end{equation}
where $S(x)$ is a scalar field and $a_{n}\ (n\ge0)$ is a countable sequence of 4-component spinor fields. Plugging the ansatz (\ref{ansatz}) into the Dirac equation (\ref{Deqn1}) and setting to zero the coefficients of the different powers of $\hbar$, we obtain
\begin{eqnarray}
\label{zeroth}
\left(\gamma^{a}S_{,a}+mI\right)a_{0} & = & 0,  \\
\label{higher}
\left(\gamma^{a}S_{,a}+mI\right)a_{n+1} & = & -\gamma^{a}a_{n;a} \ \ \ (\forall n\in\m N),
\end{eqnarray}
The existence of a solution to the homogeneous equation (\ref{zeroth}) requires
\begin{equation}
\label{hje}
\det(\gamma^{a}S_{,a}+mI)= 0.
\end{equation}
Equation (\ref{hje}) in equivalent to the Hamilton-Jacobi equation for timelike geodesics,
\begin{equation}
\label{HJ}
S^{,a}S_{,a}=m^{2}.
\end{equation}
We define
\begin{equation}
\label{4mom}
p_{a}:=-S_{,a},
\end{equation}
and normalize to obtain a unit-norm vector field 
\begin{equation}
\label{geoU}
U^{a}:=\frac1{m}p^{a} = -\frac1mS^{,a}.
\end{equation}
As a result of the Hamilton-Jacobi equation (\ref{HJ}), the integral curves of $\bb U$, defined by (\ref{geoU}), are guaranteed to be timelike geodesics.  And since equation (\ref{zeroth}) is the classical limit ($\hbar\rightarrow0$) for the Dirac equation with the semiclassical ansatz, we interpret $p_{a}$, given by equation (\ref{4mom}), as the 4-momentum, and $U^{a}$, given by (\ref{geoU}), as the 4-velocity of the Dirac particle described by $\Psi$. 

The use of the semiclassical ansatz for the Dirac equation (\ref{ansatz}) and the observation that the homogeneous equation (\ref{zeroth}) implies the Hamilton-Jacobi equation for spinless particles goes back to Pauli \cite{Pauli32}. Bargmann et al. derived equations describing the classical trajectories of spin-$\half$ particles in uniform and constant electric and magnetic fields \cite{Bargmann59}. Rubinow and Keller showed that the classical equations of Bargmann et al. could be obtained from the Dirac equation using the semiclassical ansatz \cite{Rubinow63}. Rafenelli and Schiller obtained essentially the same result soon after, using the so-called ``squared'' Dirac equation  along with the semiclassical ansatz \cite{Rafanelli64}. 

We follow the more recent work by \cite{Audretsch81} who showed that the spin vector, defined by equation (\ref{def}), is parallel propagated along the particle's trajectory to zeroth order in the asymptotic expansion (\ref{ansatz}). We will extend the result of \cite{Audretsch81} and show that, in Kerr geometry, the spin vector is parallel propagated along $U^{a}$ to first-order in $\hbar$. We begin our analysis with recalling some results from \cite{Audretsch81}. 


The matrix acting on $a_{0}$ in equation (\ref{zeroth}) is of rank 2, as is manifest by considering a frame that is parallel propagated along the congruence $U^{a}$ in which the 4-momentum takes the form $p^{(a)}=(m,0,0,0)$. The matrix acting on $a_{0}$ in equation (\ref{zeroth}) is then given by
\begin{equation}
\left(\begin{array}{cccc}0 & 0 & 0 & 0 \\0 & 0 & 0 & 0 \\0 & 0 & 2m & 0 \\0 & 0 & 0 & 2m\end{array}\right).
\end{equation}
The general solution of the homogeneous equation (\ref{zeroth}) therefore takes the form
\begin{equation}
\label{a0def}
a_{0}=\beta_{1}b_{01}+\beta_{2}b_{02},
\end{equation}
where the basis 4-spinors $b_{01}$ and $b_{02}$ are two linearly independent solutions
\begin{equation}
\label{fundamental}
b_{01}=\sqrt{\frac{E+m}{2m}}\left[\begin{array}{c}1 \\0  \\\frac{p^{\hat3}}{E+m} \\\frac{p^{\hat1}+ip^{\hat2}}{E+m}\end{array}\right], \ \ \ b_{02}=\sqrt{\frac{E+m}{2m}}\left[\begin{array}{c} 0 \\1 \\\frac{p^{\hat1}-ip^{\hat2}}{E+m}\\-\frac{p^{\hat3}}{E+m}\end{array}\right],
\end{equation}
where $E:=p^{0}$ is the energy of the Dirac particle and $p^{a}$ is defined by equation (\ref{4mom}). 
The basis spinors (\ref{fundamental}) are parallel propagated along $\bb U$,
\begin{equation}
b_{01;a}U^{a} = 0, \ \ \ b_{02;a}U^{a}=0.
\end{equation}
Choosing a frame that is parallel propagated along the congruence $U^{a}$ in which the 4-momentum takes the form $p^{(a)}=(m,0,0,0)$, the spinors $b_{01}$ and $b_{02}$ reduce to 
\begin{equation}
b_{01}=\left[\begin{array}{c}1 \\  0\\ 0 \\0 \end{array}\right], \ \ \ b_{02}= \left[\begin{array}{c} 0 \\1 \\ 0 \\0 \end{array}\right].
\end{equation}

With the inner product on spinors defined by $\langle\chi,\psi\rangle:=\bar\chi\psi=\bar\psi\chi$, for arbitrary 4-spinors $\chi,\psi$, an orthonormal basis for the 2-plane orthogonal to the fundamental solutions (\ref{fundamental}) of the homogeneous system (\ref{zeroth}) is given by the spinors
\begin{equation}
b_{11} = \sqrt{\frac{E+m}{2m}}\left[\begin{array}{c}\frac{p^{\hat3}}{E+m} \\\frac{p^{\hat1}+ip^{\hat2}}{E+m}\\1\\0 \end{array}\right], \ \ \ b_{12} =\sqrt{\frac{E+m}{2m}}\left[\begin{array}{c}\frac{p^{\hat1}-ip^{\hat2}}{E+m} \\-\frac{p^{\hat3}}{E+m}\\ 0\\1\end{array}\right].
\end{equation}
The spinor fields $(b_{01}, b_{02}, b_{11}, b_{12})$ constitute a basis for 4-spinors. In a parallel propagated frame along the congruence $U^{a}$ in which the 4-momentum takes the form $p^{(a)}=(m,0,0,0)$, $b_{11}$ and $b_{12}$ take the form
\begin{equation}
b_{11}=\left[\begin{array}{c}0 \\  0\\ 1 \\0 \end{array}\right], \ \ \ b_{12}= \left[\begin{array}{c} 0 \\0 \\ 0 \\1 \end{array}\right].
\end{equation}
These spinors are parallel propagated along $\bb U$ as well,
\begin{equation}
b_{11;a}U^{a} = 0, \ \ \ b_{12;a}U^{a}=0.
\end{equation}
The general solution to (\ref{higher}) for $n=1$ can be written as 
\begin{equation}
\label{soln1}
a_{1}=v_{1}b_{01}+v_{2}b_{02}+\lambda_{1}b_{11}+\lambda_{2}b_{12}. 
\end{equation}

We are now in a position to state the result from \cite{Audretsch81} that would allow us to derive a propagation equation for the spin vector in Kerr geometry. 

\begin{lemma}[Audretsch] The propagation equation for the scalar coefficients ($\beta=\beta_{1}, \beta_{2}, v_{1}, v_{2}$) of the fundamental spinors $b_{01}$ and $b_{02}$ in the general solution (\ref{soln1}) to the first-order equation (\ref{higher}) with $n=1$ is given by
\begin{equation}
\label{propeqn} 
\beta_{,a}U^{a} = -\frac\theta2 \beta. 
\end{equation}
\end{lemma}
In order to prove that the spin vector defined by (\ref{def}) is parallel propagated to first-order in the asymptotic expansion (\ref{ansatz}), we will now to determine the solutions of the propagation equation for the scalar coefficients in Kerr geometry. 
\begin{theorem}[Scalar coefficients in Kerr geometry]\label{scalarcoeff}
Let $\bb U$ be tangent to a timelike geodesic in Kerr geometry. The general solution to the propagation equation for the scalar coefficients (\ref{propeqn}) is given by
\begin{equation}
\label{gensoln}
\beta(r,\vartheta):=\frac{c}{\sqrt{R(r)\Theta(\vartheta)\sin\vartheta}},
\end{equation}
where $c$ is a constant of integration. 
\begin{proof}
We may rewrite the propagation equation (\ref{propeqn}) in terms of proper time $\tau$ as follows.
\begin{eqnarray}
\frac{d}{d\tau}\ln\beta(\tau)&=&-\frac12\theta(\tau),\\
\label{scalarproptime}
 & = & -\frac12\left(\frac{R'}{R}\dot r+\frac{\Theta'}{\Theta}\dot\vartheta+\cot\vartheta\dot\vartheta\right),\\
 \label{kerrscprop}
 & = & -\frac12\frac{d}{d\tau}\ln(R\Theta\sin\vartheta),
\end{eqnarray}
where the dot denotes $\frac{d}{d\tau}$, prime denotes ordinary partial derivatives, and where the functions $R(r)$ and $\Theta(\vartheta)$ are defined by equations (\ref{defR}) and (\ref{defTheta}) respectively. It follows from equation (\ref{kerrscprop}) that the general solution to (\ref{propeqn}) is given by the scalar field (\ref{gensoln}).
\end{proof}
\end{theorem}
\noindent Let
\begin{eqnarray}
\label{scal1}
\beta_{1} & = & \frac{c_{1}}{\sqrt{R(r)\Theta(\vartheta)\sin\vartheta}}, \\
\beta_{2} & = & \frac{c_{2}}{\sqrt{R(r)\Theta(\vartheta)\sin\vartheta}}, \\
v_{1} & = & \frac{d_{1}}{\sqrt{R(r)\Theta(\vartheta)\sin\vartheta}}, \\
\label{scal2}
v_{2} & = & \frac{d_{2}}{\sqrt{R(r)\Theta(\vartheta)\sin\vartheta}}. 
\end{eqnarray}
We can easily verify that 
\begin{equation}
\label{fdef}
f:=\bar a_{0}a_{0}=\sqrt{\beta^{*}_{1}\beta_{1}+\beta^{*}_{2}\beta_{2}},
\end{equation}
satisfies the propagation equation
\begin{equation}
\label{fprop}
f_{,a}U^{a}= -\frac\theta2. 
\end{equation}
Define a spinor $b_{0}$ such that 
\begin{equation}
\label{b0def}
a_{0}=fb_{0}.
\end{equation}
Then, $b_{0}$ has unit norm,
\begin{equation}
\bar b_{0}b_{0}=1,
\end{equation}
and is parallel propagated along $\bb U$,
\begin{equation}
b_{0;a}U^{a}=0.
\end{equation} 
We now recall our definition of the spin vector given by equation (\ref{def}), 
\begin{equation}
W^{a} := \frac12\varepsilon^{abcd}J_{b}\frac{\bar\Psi\sigma_{cd}\Psi}{\bar\Psi\Psi},
\end{equation}
where $\varepsilon^{abcd}$ is the Levi-Civita symbol. In order to derive a propagation equation for the spin vector to first-order in $\hbar$, we note that
\begin{equation}
\frac1{\bar\Psi\Psi}=\frac1{f^{2}}\left(1+\frac{i\hbar}{f^{2}}\left(\bar a_{0}a_{1}-\bar a_{1}a_{0}\right)\right)+O(\hbar^{2}).
\end{equation}
Therefore, the spin vector is given by
\begin{equation}
\label{w}
W^{a}=W_{0}^{a}+\hbar W^a_{1}+O(\hbar^{2}), 
\end{equation}
where
\begin{equation}
\label{w0}
W_{0}^{a}= \frac12\varepsilon^{abcd}U_{b}\bar b_{0}\Sigma_{cd}b_{0},
\end{equation}
and
\begin{align}
\label{w1}
\nonumber
W_{1}^{a} = &\frac{i}2\varepsilon^{abcd}\bigg[U_{b}\left(\frac{\bar a_{0}a_{1}-\bar a_{1}a_{0}}{f^{2}}\bar b_{0}\sigma_{cd}b_{0}-\frac1{f^{2}}\left(\bar a_{0}\sigma_{cd}a_{1}-\bar a_{1}\sigma_{cd}a_{0}\right)\right)\\
& -\frac{1}{2m}\left(\bar b_{0;b}b_{0}-\bar b_{0}b_{0;b}\right)\bar b_{0}\sigma_{cd}b_{0}\bigg]. 
\end{align}

The spin vector defined by (\ref{def}) and given by (\ref{w}) satisfies, 
\begin{equation}
W^{a}W_{a} = -1 + O(\hbar^{2}). 
\end{equation}  
In order to prove that the spin vector defined by (\ref{def}) is parallel propagated to first-order in the asymptotic expansion (\ref{ansatz}) in Kerr geometry we will show that, to first-order in $\hbar$, the spin vector defined by (\ref{def}) has constant components in the parallel propagated frame. 



\section{Propagation law for the spin vector}\label{proplaw}
In order to prove that the spin vector defined by (\ref{def}) is parallel propagated to first-order in the formal asymptotic expansion (\ref{ansatz}) in Kerr geometry, we will need the following lemma. 

\begin{lemma}
For the general solution to the propagation equation for the scalar coefficients in Kerr geometry $\beta(r,\vartheta)$, given by (\ref{gensoln}), we have
\begin{equation}
\beta_{,(0)} =-\frac\theta2\beta,
\end{equation}
\begin{equation}
\label{betacomma1}
\beta_{,(1)}=-\frac\beta{2\Sigma}\left[\varpi R'\left(\cos\Phi r\m P\kappa^{-\half}-\sin\Phi\right)+\frac{\cot\vartheta+\Theta'/\Theta}\varpi\left(\cos\Phi a\cos\vartheta\m D\kappa^{-\half} -\sin\Phi\Theta\right)\right],
\end{equation}
\begin{equation}
\label{betacomma2}
\beta_{,(2)}=-\frac\beta{2\Sigma}\left[\varpi R'\left(\sin\Phi r\m P\kappa^{-\half}+\cos\Phi\right)+\frac{\cot\vartheta+\Theta'/\Theta}\varpi\left(\sin\Phi a\cos\vartheta\m D\kappa^{-\half} +\cos\Phi\Theta\right)\right],
\end{equation}
\begin{equation}
\label{betacomma3}
\beta_{,(3)}=-\frac{\beta}{2\kappa^{\half}\Sigma}\left[-a\cos\vartheta\m P \frac{R'}{R} + r\m D\left(\cot\vartheta+\frac{\Theta'}{\Theta}\right)\right],
\end{equation}
where the prime denotes ordinary derivatives. 
\begin{proof}
Let $\langle X,Y\rangle:=X^{a}Y_{a}$ denote the inner product for arbitrary vector fields $X$ and $Y$. Observe that $\beta_{,(a)}=\langle d\beta,L_{(a)}\rangle$. The proof then follows from direct computation using the parallel propagated frame $L_{(a)}$ constructed in Section \ref{FFF}. 
\begin{equation}
\langle d\beta,L_{(0)}\rangle=\langle d\beta,U\rangle= -\frac\theta2\beta.
\end{equation}
\begin{equation}
\label{dbeta1}
\langle d\beta,\widetilde L_{(1)}\rangle =-\frac{\beta}{2\kappa^{\half}\Sigma}\left[r\varpi\m P \frac{R'}{R} + \frac{a\cos\vartheta}\varpi\m D\left(\cot\vartheta+\frac{\Theta'}{\Theta}\right)\right],
\end{equation}
\begin{equation}
\label{dbeta2}
\langle d\beta,\widetilde L_{(2)}\rangle =-\frac{\beta}{2\Sigma}\left[\varpi R' +\frac{\Theta}\varpi\left(\cot\vartheta+\frac{\Theta'}\Theta\right)\right],
\end{equation}
\begin{equation}
\langle d\beta, L_{(3)}\rangle =-\frac{\beta}{2\kappa^{\half}\Sigma}\left[-a\cos\vartheta\m P \frac{R'}{R} + r\m D\left(\cot\vartheta+\frac{\Theta'}{\Theta}\right)\right].
\end{equation}
\end{proof}
\end{lemma}

\begin{theorem}\label{ppo1}
In Kerr geometry, to first-order in $\hbar$, the spin vector defined by (\ref{def}) is parallel propagated along $\bb U$.
\begin{proof}
Theorem \ref{scalarcoeff} guarantees that the scalar coefficients of the spinor fields take the form (\ref{scal1})-(\ref{scal2}) for some constants of integration $c_{1},c_{1},d_{1},d_{2}$. Plugging the scalar coefficients  (\ref{scal1})-(\ref{scal2}) into the expressions previously established for the spin vector, (\ref{w0}) and (\ref{w1}), we obtain in the parallel propagated frame
\begin{equation}
\label{ }
W_{0}=\frac1{c_{1}^{*}c_{1}+c_{2}^{*}c_{2}}\left[\begin{array}{c}0\\c_{1}^{*}c_{2}+c_{2}^{*}c_{1} \\i(c_{1}^{*}c_{2}-c_{2}^{*}c_{1}) \\c_{2}^{*}c_{2}-c_{1}^{*}c_{1}\end{array}\right],
\end{equation}
and
\begin{equation}
W_{1}=P+Q+R,
\end{equation}
where
\begin{equation}
P=\frac{i}{f^{2}}\left(\bar a_{0}a_{1}-\bar a_{1}a_{0}\right)W_{0}=i\frac{c_{1}^{*}d_{1}+c_{2}^{*}d_{2}-d_{1}^{*}c_{1}-d_{2}^{*}c_{2}}{c_{1}^{*}c_{1}+c_{2}^{*}c_{2}}W_{0},
\end{equation}
\begin{eqnarray}
Q&=&\frac{i}{2f^{2}}\varepsilon^{(a)(0)\hat i\hat j}\left(\bar a_{1}\sigma_{\hat i\hat j}a_{0}-\bar a_{0}\sigma_{\hat i\hat j}a_{1}\right),\\
&=&\frac{i}{c_{1}^{*}c_{1}+c_{2}^{*}c_{2}}\left[\begin{array}{c}0\\\left(d_{1}^{*}c_{2}+d_{2}^{*}c_{1}\right)-\left(c_{1}^{*}d_{2}+c_{2}^{*}d_{1}\right) \\ i\left(\left(c_{2}^{*}d_{1}+d_{1}^{*}c_{2}\right)-\left(c_{1}^{*}d_{2}+d_{2}^{*}c_{1}\right)\right) \\\left(c_{1}^{*}d_{1}+d_{2}^{*}c_{2}\right)-\left(d_{1}^{*}c_{1}+c_{2}^{*}d_{2}\right)\end{array}\right],
\end{eqnarray}
and
\begin{equation}
\label{third}
R=\frac12\varepsilon^{(a)(b)(c)(d)}\delta J_{(b)}\bar b_{0}\sigma_{(c)(d)}b_{0},
\end{equation}
and where $\delta J$ is given by
\begin{equation}
\label{vcorr}
\delta J_{(a)}=\frac{1}{2mi}\left(\bar b_{0;(a)}b_{0}-\bar b_{0}b_{0;(a)}\right).
\end{equation}
\begin{remark}
Note that $P$ and $Q$ have constant components in a parallel propagated frame. We will now show that $R$, given by (\ref{third}), vanishes identically. 
\end{remark}
\noindent We begin by evaluating (\ref{vcorr}).\footnote{Recall that $a_{0}=fb_{0}=\beta_{1}b_{01}+\beta_{2}b_{02}$ and $f=\sqrt{\beta_{1}^{*}\beta_{1}+\beta_{2}^{*}\beta_{2}}$ in accordance with (\ref{fdef}), (\ref{b0def}), and (\ref{a0def}).}
\begin{equation}
\label{ }
\bar b_{0;(a)}b_{0}-\bar b_{0}b_{0;(a)}=\frac{\left(\beta_{1}L_{(a)}\left(\beta_{1}^{*}\right)+\beta_{2}L_{(a)}\left(\beta_{2}^{*}\right)\right)-\left(\beta_{1}^{*}L_{(a)}\left(\beta_{1}\right)+\beta_{2}^{*}L_{(a)}\left(\beta_{2}\right)\right)}{f^{2}},
\end{equation}
where $L_{(a)}(\beta):=\langle d\beta,L_{(a)}\rangle$. Since $L_{(0)}(\beta)=\langle d\beta,U\rangle=-\frac\theta2\beta$, we have
\begin{equation}
\label{ }
\bar b_{0;(0)}b_{0}-\bar b_{0}b_{0;(0)}=0. 
\end{equation}
And for the spacelike basis vectors of the parallel propagated frame $L_{(i)}$,
\begin{equation}
\label{ }
\bar b_{0;(i)}b_{0}-\bar b_{0}b_{0;(i)}=\frac{\beta_{1}\beta_{1,(i)}^{*}+\beta_{2}\beta_{2,(i)}^{*}-\beta_{1}^{*}\beta_{1,(i)}-\beta_{2}^{*}\beta_{2,(i)}}{f^{2}}
\end{equation}
Since 
\begin{equation}
\label{spcomp3}
\bar b_{0}\sigma_{(a)(0)}b_{0}=0 \ \ \textup{and} \ \ b_{0;(0)}=0,
\end{equation}
at least one of the three terms, $\varepsilon^{(i)(a)(b)(c)}$,  $\delta J_{(a)}$, and $\bar b_{0}\sigma_{(b)(c)}b_{0}$, vanishes for any choice of indices $(a),(b),(c)$, one of which has to be $(0)$. Thus, $R^{(i)}$ must vanish for $i=1,2,3$. 

\noindent The timelike component of $R$ is given by
\begin{align}
\label{three}
\nonumber
R^{(0)}  & =\frac1{2mif^{4}}\bigg[\left(\beta_{1}^{*}\beta_{2}+\beta_{2}^{*}\beta_{1}\right)\left(\beta_{1,(1)}\beta_{1}^{*}-\beta_{1,(1)}^{*}\beta_{1}+\beta_{2,(1)}\beta_{2}^{*}-\beta^{*}_{2,(1)}\beta_{2}\right)  \\
 \nonumber
    &  +i\left(\beta_{1}^{*}\beta_{2}-\beta_{2}^{*}\beta_{1}\right)\left(\beta_{1,(2)}\beta_{1}^{*}-\beta_{1,(2)}^{*}\beta_{1}+\beta_{2,(2)}\beta_{2}^{*}-\beta^{*}_{2,(2)}\beta_{2}\right) \\
    &  +\left(\beta_{2}^{*}\beta_{2}-\beta_{1}^{*}\beta_{1}\right)\left(\beta_{1,(3)}\beta_{1}^{*}-\beta_{1,(3)}^{*}\beta_{1}+\beta_{2,(3)}\beta_{2}^{*}-\beta^{*}_{2,(3)}\beta_{2}\right)\bigg].
\end{align}
Note that we have a common factor between the derivatives of the scalar fields (\ref{betacomma1})-(\ref{betacomma3}). Define
\begin{equation}
\zeta(i):=\frac{\beta_{1,(i)}}{\beta_{1}}=\frac{\beta_{2,(i)}}{\beta_{2}}=\frac{\beta^{*}_{1,(i)}}{\beta^{*}_{1}}=\frac{\beta^{*}_{2,(i)}}{\beta_{2}^{*}}.
\end{equation}
Therefore we may replace each term, $\beta_{k,(i)}$, by $\zeta(i)\beta_{k}$  (for $k=1,2,3$) in the expression for $R^{(0)}$ given by  (\ref{three}) as follows. 
\begin{align}
\label{}
\nonumber
R^{(0)}  & =\frac1{2mif^{4}}\bigg[\zeta(1)\left(\beta_{1}^{*}\beta_{2}+\beta_{2}^{*}\beta_{1}\right)\left(\beta_{1}\beta_{1}^{*}-\beta_{1}^{*}\beta_{1}+\beta_{2}\beta_{2}^{*}-\beta^{*}_{2}\beta_{2}\right)  \\ 
 \nonumber
    &  +i\zeta(2)\left(\beta_{1}^{*}\beta_{2}-\beta_{2}^{*}\beta_{1}\right)\left(\beta_{1}\beta_{1}^{*}-\beta_{1}^{*}\beta_{1}+\beta_{2}\beta_{2}^{*}-\beta^{*}_{2}\beta_{2}\right) \\   
    &  +\zeta(3)\left(\beta_{2}^{*}\beta_{2}-\beta_{1}^{*}\beta_{1}\right)\left(\beta_{1}\beta_{1}^{*}-\beta_{1}^{*}\beta_{1}+\beta_{2}\beta_{2}^{*}-\beta^{*}_{2}\beta_{2}\right)\bigg],\\
    \nonumber
    & = 0.
\end{align}
This completes the proof of our claim that $R$ vanishes identically. 

Since ($c_{1}, c_{2}, d_{1},d_{2}$) are constant along the integral curve of $\bb U$, we see that, to first order in $\hbar$, the spin vector has constant components in the parallel propagated frame, and is thus parallel propagated along $\bb U$.
\end{proof}
\end{theorem}
\noindent For the sake of completeness, we note the expression of the spin vector for a Dirac particle in Kerr geometry to first-order in $\hbar$.
\begin{equation}
W^{a}=W_{0}^{a}+\hbar W_{1}^{a}+O(\hbar^{2}),
\end{equation}
where
\begin{equation}
\label{ }
W_{0}=\frac1{c_{1}^{*}c_{1}+c_{2}^{*}c_{2}}\left[\begin{array}{c}0\\c_{1}^{*}c_{2}+c_{2}^{*}c_{1} \\i(c_{1}^{*}c_{2}-c_{2}^{*}c_{1}) \\c_{2}^{*}c_{2}-c_{1}^{*}c_{1}\end{array}\right],
\end{equation}
and
\begin{equation}
\label{ }
W_{1}=\frac{i}{c_{1}^{*}c_{1}+c_{2}^{*}c_{2}}\left(c_{0}W_{0}+\left[\begin{array}{c}0\\\left(d_{1}^{*}c_{2}+d_{2}^{*}c_{1}\right)-\left(c_{1}^{*}d_{2}+c_{2}^{*}d_{1}\right) \\ i\left(\left(c_{2}^{*}d_{1}+d_{1}^{*}c_{2}\right)-\left(c_{1}^{*}d_{2}+d_{2}^{*}c_{1}\right)\right) \\\left(c_{1}^{*}d_{1}+d_{2}^{*}c_{2}\right)-\left(d_{1}^{*}c_{1}+c_{2}^{*}d_{2}\right)\end{array}\right]\right),
\end{equation}
where $c_{0}:=(c_{1}^{*}d_{1}+c_{2}^{*}d_{2})-(d_{1}^{*}c_{1}+d_{2}^{*}c_{2})$. 

\section{Reference frame for Dirac particles}\label{measure}
\noindent  In this section, we will construct a measurement frame on purely geometric criteria and specify the communication protocol. We seek a construction that is coordinate independent and as closely tied to the symmetries of Kerr geometry as possible. In the photonic case, one could use the principal null directions of the Weyl tensor to define a pair of basis vectors for the plane of polarization. Unfortunately, that construction cannot serve us in the case of massive spin-$\half$ particles since the spin vector cannot in general be confined to a 2-plane, as was the case with the polarization vector of a photon. 

Recall that $\bb U$ denotes the 4-velocity of the Dirac particle. We first define a volume form for the 3-space $\langle\bb U\rangle^{\perp}$ by
\begin{equation}
\label{3form}
\bb\Omega:=\star\bb U^{\flat},
\end{equation}
where $\star$ denotes the Hodge duality operator. Since 
\begin{equation}
\label{ }
\bb U^{\flat}:=\left(\Delta\Sigma\right)^{-\half}\left(\m P\bb\omega^{0}-R^{\half}\bb\omega^{1}-\Delta^{\half}\m D\bb\omega^{2}-\Delta^{\half}\bb\omega^{3}\right),
\end{equation}
the volume form $\bb\Omega$, defined by (\ref{3form}), is given
\begin{eqnarray}
\nonumber
\left(\Delta\Sigma\right)^{\half}\bb\Omega &=&-\Delta^{\half}\Theta^{\half}\bb\omega^{0}\bb\wedge\bb\omega^{1}\wedge\bb\omega^{2}+\Delta^{\half}\m D\bb\omega^{0}\wedge\bb\omega^{1}\wedge\bb\omega^{3}\\
\label{Omega}
&&-R^{\half}\bb\omega^{0}\wedge\bb\omega^{2}\wedge\bb\omega^{3}+\m P \bb\omega^{1}\wedge\bb\omega^{2}\wedge\bb\omega^{3}.
\end{eqnarray} 
The volume form $\bb\Omega$ given by (\ref{Omega}) should be a monomial. This is indeed the case if we consider 1-forms restricted to $\langle U\rangle^{\perp}$. Let $\bb\alpha^{\hat i}:=\bb L_{(\hat i)}^{\flat}$ for $\hat i=\hat1,\hat2,\hat3$, where $\bb L_{(a)}$ is the parallel propagated frame constructed in Section \ref{Kerr}. Then, it follows from (\ref{Omega}) that 
\begin{equation}
\bb\Omega=\bb\alpha^{\hat 1}\wedge\bb\alpha^{\hat 2}\wedge\bb\alpha^{\hat 3}. 
\end{equation} 

Consider the symmetric frame dual to the coframe defined by (\ref{carter}). Since one of the basis 4-vectors, ($\bb e_{0}$), is timelike and three ($\bb e_{i}, i=1,2,3$) are spacelike, there is still an ambiguity in labeling the spacelike basis 4-vectors corresponding to six permutations. We show how Carter observers can use geometric criteria to eliminate this ambiguity and agree on the labeling of the symmetric frame. 

The symmetric linear combination of the principal null directions of the Weyl tensor, $2^{-\half}\left(\bb\ell+\bb n\right)$, is timelike, while their difference $2^{-\half}\left(\bb\ell-\bb n\right)$ is spacelike. This is a geometrically privileged vector that we can label $\bb e_{1}$. Since the observers' 4-velocity vector $\bb e_{0}:=2^{-\half}\left(\bb\ell+\bb n\right)$ is known, they can identify their spacelike 4-acceleration vector, whose components in the symmetric frame are given by
\begin{equation}
\label{acceleration}
\nabla_{\bb e_{0}}\bb e_{0}= \Sigma^{-\frac32}\left(0,\Delta^{\half}\left((r^{2}-a^{2}\cos^{2}\vartheta)M-ra^{2}\sin^{2}\vartheta\right), 0, -a^{2}\cos\vartheta\sin\vartheta\right).
\end{equation}
That is, of the two remaining vectors of the symmetric frame to be labeled, the acceleration vector, given by (\ref{acceleration}), is orthogonal to only one of them, which we label $\bb e_{2}$. The last remaining basis 4-vector is labeled $\bb e_{3}$. Thus, there is no ambiguity in labeling the indices of the symmetric frame. 

We now obtain a basis for $\langle \bb U\rangle^{\perp}$ by contracting the spacelike basis 4-vectors with $\bb\Omega$ in an increasing sequence ($i<j$) and performing a Gram-Schmidt orthonormalization. 
One obtains
\begin{eqnarray}
\label{meas1}
\bb X&:=&\rho^{-\half}\left(-\Delta^{\half}\Theta^{\half}, 0,0,-\m P\right),\\
\label{meas2}
\bb Y&:=&\rho^{-\half}\varrho^{-\half}\left(\Delta^{\half}\m P\m D, 0,\rho, \Delta\m D\Theta^{\half} \right),\\
\label{meas3}
\bb Z&:=&\left(\Delta\Sigma\varrho\right)^{-\half}\left(-\m PR^{\half}, -\varrho, -\m D\Delta^{\half}R^{\half}, -\Delta^{\half}\Theta^{\half}R^{\half}\right),
\end{eqnarray}
where
\begin{equation}
\label{varrho}
\varrho:=\m P^{2}-\Delta(\kappa-a^{2}\cos^{2}\vartheta), \ \ \rho := \m P^{2}-\Delta\Theta,
\end{equation}
 
\begin{definition}[Reference 3-frame]\label{ref3def}
The set of three spacelike 4-vectors, $\{\bb X,\bb Y, \bb Z\}$, given by (\ref{meas1})-(\ref{meas3}), constitutes an orthonormal basis for $\langle \bb U\rangle^{\perp}$, that we call the \emph{reference 3-frame}. 
\end{definition}
We are now in a position to specify our communication protocol. Let Alice and Bob be two Carter observers located in Kerr geometry. In order to communicate with Bob, Alice polarizes a massive spin-$\half$ at event $x_{A}$ by choosing a unit-norm 3 component vector $\whector W$ in the reference 3-frame $\{\bb X,\bb Y, \bb Z\}$ at event $x_{A}$, and launches it on a timelike geodesic $\gamma(\tau)$ that intersects with Bob's worldline. Bob intercepts the particle and measures its spin vector by projecting it onto the reference 3-frame $\{\bb X,\bb Y, \bb Z\}$ at event $x_{B}$. The timelike geodesic $\gamma$ must satisfy
\begin{equation}
\gamma(\tau_{0}) = x_{A}, \ \ \ \text{and} \ \ \  \gamma(\tau_{1}) = x_{B}, 
\end{equation} 
for some $\tau_{1}>\tau_{0}$. We may without loss of generality set $\tau_{0}=0$, and supress the subscript for $\tau_{1}$. Thus, $\gamma(0)=x_{A}$ and $\gamma(\tau)=x_{B}$. 

Armed with the definitions of the reference 3-frame and the communication protocol, we can define the geometrically-induced precession of a massive spin-$\half$ particle as follows. 
\begin{definition}[Geometrically-induced precession of the spin vector]\label{DiracPrecession}
The precession of the spin of the Dirac particle is given by the proper time dependent rotation $\Lambda(\tau)\in SO(3)$ of the reference 3-frame $\{\bb X,\bb Y, \bb Z\}$ with respect to the 3-frame $L_{(i)}$ that is parallel propagated along $\bb U$ and also spans $\langle\bb U\rangle^{\perp}$. More precisely, we have 
\begin{equation}
\label{rotdef}
\Lambda^{\hat i}_{\phantom{\hat i}\hat j}(\tau) := L_{\phantom{\hat i}(k)}^{\hat i}(\tau) L ^{(k)}_{\phantom{(k)}\hat j}(0),
\end{equation}
where $L ^{(k)}_{\phantom{(k)}\hat j}(0)$ is the change-of-basis matrix from the reference 3-frame to the parallel propagated 3-frame $L_{(k)}(0)$ at event $x_{A}$, whereas $L_{\phantom{\hat i}(k)}^{\hat i}(\tau)$ is the change-of-basis matrix from the parallel propagated 3-frame $L_{(k)}(\tau)$ to the reference 3-frame at event $x_{B}$. 
\end{definition}
\begin{remark}
In equation (\ref{rotdef}), the hatted indices refer to the reference 3-frame and the bracketed indices refer to the parallel propagated 3-frame. We are thus identifying the tangent spaces along $\gamma$ using the parallel propagated frame. Under this identification of tangent spaces, the matrix $\Lambda^{\hat i}_{\phantom{\hat i}\hat j}(\tau)$, defined by (\ref{rotdef}), rotates Alice's reference 3-frame to Bob's reference 3-frame. Thus, 
\begin{equation}
\whector W_{B}=\Lambda(\tau)\whector W_{A}.
\end{equation}
\end{remark}

In the next section, we obtain an exact expression for the rotation $\Lambda(\tau)$, and obtain a curvature invariant for the curve traced out by the spin vector on the unit sphere under the action of $\Lambda(\tau)$. 

\section{Geometrically-induced rotation of the spin vector}\label{precession}
\noindent Direct computation yields
\begin{equation}
\label{rotexpr}
\Lambda(\tau):=\left(\begin{array}{ccc}\frac{r\m D\m P+a\cos\vartheta\sqrt{R\Theta}}{\sqrt{\kappa\Sigma\rho}} & -\frac{\kappa^{\half}\sin\chi\Sigma\m P\Theta^{\half}+c_{1}\cos\chi}{\varpi^{\half}(\kappa+r^{2})\sqrt{\kappa\Sigma\rho}} & \frac{\kappa^{\half}\cos\chi\Sigma\m P\Theta^{\half}-c_{1}\sin\chi}{\varpi^{\half}(\kappa+r^{2})\sqrt{\kappa\Sigma\rho}} \\ \frac{r\varrho\Theta^{\half}-a\cos\vartheta\m D\m PR^{\half}}{\sqrt{\kappa\Sigma\varrho\rho}} & \frac{\kappa^{\half}\sin\chi\Sigma\m D\m P^{2}+c_{2}\cos\chi}{\varpi^{\half}(\kappa+r^{2})\sqrt{\kappa\Sigma\varrho\rho}} & \frac{-\kappa^{\half}\cos\chi\Sigma\m D\m P^{2}+c_{2}\sin\chi}{\varpi^{\half}(\kappa+r^{2})\sqrt{\kappa\Sigma\varrho\rho}} \\-\frac{a\cos\vartheta\m P}{\sqrt{\kappa\varrho}} & \varpi^{\half}\frac{\cos\chi r\m P-\kappa^{\half}\sin\chi R^{\half}}{\sqrt{\kappa\varrho}} & \varpi^{\half}\frac{\sin\chi r\m P+\kappa^{\half}\cos\chi R^{\half}}{\sqrt{\kappa\varrho}}\end{array}\right),
\end{equation}
where
\begin{equation}
c_{1}:=(\kappa+r^{2})\left(r\varpi\sqrt{R\Theta}-a\cos\vartheta\m D\m P\right), \ \ c_{2}:=(\kappa+r^{2})\left(a\cos\vartheta\varrho\Theta^{\half}+r\varpi\m D\m PR^{\half}\right), 
\end{equation}
and $\varpi$ is defined by equation (\ref{varpi}), while $\varrho$ and $\rho$ were defined in equation (\ref{varrho}). 
In the matrix on the RHS of equation (\ref{rotexpr}), all entries depends on proper time $\tau$ through $r(\tau)$ and $\vartheta(\tau)$. We verify that $\Lambda^{T}\Lambda=I$ and $\det\Lambda=1$. Thus, $\Lambda\in SO(3)$.

Let $\whector W_{0}$ be the initial position of the spin vector prepared by Alice. At proper time $\tau>0$, the position vector is given by 
\begin{equation}
\label{posvec}
\whector W(\tau) = \Lambda(\tau)\whector W_{0},
\end{equation}
where $\Lambda(\tau)$ is given by (\ref{rotexpr}). 

The differential equation satisfied by the Darboux frame $\left(\whector W, \whector W', \whector W\times\whector W'\right)$ for a curve on the unit sphere can be written as \cite{Guggenheimer77}
\begin{equation}
\label{Darboux}
\left(\begin{array}{c}\whector W \\ \whector W' \\ \whector W\times\whector W'\end{array}\right)'=\left(\begin{array}{ccc}0 & 1 & 0 \\-1 & 0 & k_{g} \\0 & -k_{g} & 0\end{array}\right)\left(\begin{array}{c}\whector W \\ \whector W' \\ \whector W\times\whector W'\end{array}\right),
\end{equation}
where the prime denotes derivatives taken with respect to the arclength parameter $s$ and $k_{g}(s)$ is the \emph{spherical curvature}.\footnote{The spherical curvature $k_{g}$ is usually called geodesic curvature. But we shall not that terminology to avoid confusion between geodesics on the sphere and geodesics of Kerr geometry.} Any two curves confined to the unit sphere with the same spherical curvature are related to each other by a constant rotation and conversely.\footnote{More precisely, if $\left(\alpha(s)\right)_{s\in(a,b)}$ and $\left(\beta(s)\right)_{s\in(a,b)}$ are two curves on the unit sphere parametrized by arc length $s$, then there exists  $\Lambda\in SO(3)$ such that $\alpha(s)=\Lambda\beta(s),\forall s\in(a,b)$ if and only if $k^{\alpha}_{g}(s)=k^{\beta}_{g}(s),\forall s\in(a,b)$.} 

From equation (\ref{Darboux}), we obtain the following straightforward formula for the spherical curvature of curves on the unit sphere parametrized by the arclength $s$.
\begin{equation}
\label{curvaturearclength}
k_{g}(s)=\frac{\whector W'\cdot \left(\whector W''\times\whector W\right)}{\whector W'\cdot\whector W'}.
\end{equation}
Note that our position vector (\ref{posvec}) is parametrized by proper time, $\tau$, \emph{not} by arclength $s$ of the curve traced out by the position vector $\whector W$. We therefore amend equation (\ref{curvaturearclength}) to obtain the following formula for the spherical curvature invariant in terms of proper time $\tau$. 
\begin{equation}
\label{sphericalcurvature}
k_{g}(\tau)=\frac{\frac{d}{d\tau}\whector W\cdot\left(\frac{d^{2}}{d\tau^{2}}\whector W\times\whector W\right)}{\Vert\frac{d}{d\tau}\whector W\Vert^{3}},
\end{equation}
where $\Vert\cdot\Vert$ denotes the Euclidean norm. The spherical curvature $k_{g}$ defined by the differential equation (\ref{Darboux}) and given by formula (\ref{sphericalcurvature}) is independent of the choice of initial position vector $\whector W_{0}$. Another choice of initial position vector merely induces a constant rotation of the curve that leaves the spherical curvature $k_{g}$ invariant. We are thus free to choose the initial position vector so as to minimize computational complexity. We fix the initial position vector at $\tau=0$ to be given in the reference frame (\ref{meas1})-(\ref{meas3}) by
\begin{equation}
\whector W_{0} = \left(\begin{array}{c}1 \\0 \\0\end{array}\right). 
\end{equation}
At proper time $\tau>0$, the spin vector is then given simply by 
\begin{equation}
\label{measpin}
\whector W(\tau) = \Lambda(\tau)\whector W_{0} = \left(\begin{array}{c}\frac{r\m D\m P+a\cos\vartheta\sqrt{R\Theta}}{\sqrt{\kappa\Sigma\rho}}  \\  \frac{r\varrho\sqrt\Theta-a\cos\vartheta\m D\m P\sqrt{R}}{\sqrt{\kappa\Sigma\varrho\rho}} \\ -\frac{a\cos\vartheta\m P}{\sqrt{\kappa\varrho}}\end{array}\right).
\end{equation}
We are now in a position to prove the following result. 
\begin{proposition}\label{DiracEquator}
There is no spin precession for Dirac particles confined to the equatorial plane,  $Eq:=\{-\infty<t<+\infty,r_{+}<r<+\infty, \vartheta=\pi/2, 0\le\phi<2\pi\}$, of Kerr geometry.
\begin{proof}
For timelike geodesics confined to the equatorial plane, we have $\vartheta=\frac\pi2$, $\varrho=\m P^{2}-\kappa\Delta$, $\varpi=\frac{\kappa}{r^{2}+\kappa}$ and  $\kappa=\left(aE-\Phi\right)^{2}$. Thus, 
\begin{equation}
\whector W(\tau) = \left(\begin{array}{c}1 \\0 \\0\end{array}\right),
\end{equation}
for arbitrary proper time $\tau$. 
\end{proof}
\end{proposition}
\begin{cor}\label{DiracSchwarzschild}
There is no spin precession for Dirac particles in Schwarzschild geometry. 
\begin{proof}
  Since Schwarzschild geometry is spherically symmetric, geodesics are
  confined to planes through the origin \cite{Chandrasekhar92}. Any plane through the origin in Schwarzschild geometry is isometric to the equatorial plane of a degenerate Kerr solution with $a=0$.
\end{proof}
\end{cor}
In Section \ref{discuss}, we will discuss these results as well as present some plots for qualitative analysis. 

\section{Summary and discussion}\label{discuss}
In order to isolate the geometrically-induced precession of the spin of Dirac particles we started with the Dirac equation. In general, there is no way to associate a solution of the Dirac equation (\ref{Deqn1}) to a classical trajectory; that is, a timelike geodesic along which the nonzero rest mass, spin-$\half$ test particle propagates. Following the insight of \cite{Pauli32}, we deploy the semiclassical ansatz for the Dirac spinor. The classical limit ($\hbar\rightarrow0$) for the Dirac equation with the semiclassical ansatz is equivalent to the Hamilton-Jacobi equation for spinless particles. This yields the desired geodesic that we postulate is the classical trajectory of the Dirac particle. 

In order to define the spin vector, we appealed to the decomposition of the Dirac current due to Gordon (Theorem \ref{GordonDecomp}). The spin vector defined in terms of the conserved polarization current in the Gordon decomposition (Definition \ref{spinvectordef}), corresponds to the Foldy-Wouthuysen mean-spin operator \cite{Kirsch01}. Following the discussion in Section \ref{DE}, we can be confident that this definition of the spin vector, first proposed in  \cite{Audretsch81}, is a good one. 

Extending the result of \cite{Audretsch81}, we showed that the spin vector defined in terms of the conserved polarization current in the Gordon decomposition, is parallel propagated along the geodesic obtained from the semiclassical ansatz to $O(\hbar^{2})$ in Kerr geometry (Theorem \ref{proplaw}). We were somewhat surprised by the result as we were expecting a more involved propagation law for Dirac particles. 

We showed how observers located in the outer geometry of Kerr black holes may set up reference 3-frames in terms of locally-available, purely geometric information (Definition \ref{ref3def}). We were thus able to define the geometrically-induced precession of the spin vector of Dirac particles propagating in the outer geometry of Kerr black holes (Definition \ref{DiracPrecession}). Our geometrically-motivated strategy allowed us to obtain a compact expression for the geometrically-induced spin precession for Dirac particles. We showed how the geometrically-induced precession of the spin of Dirac particles is determined by the rotation of the parallel-propagated frame with respect to the reference frame that the observers use to measure the spin of the test particles. We further showed that the geometrically-induced spin precession of a Dirac particle can be invariantly represented by the spherical curvature of the curve traced out by the spin vector on the unit sphere.

We showed how there is no spin precession for Dirac particles confined to the equatorial plane of Kerr geometry (Proposition \ref{DiracEquator}). The significance of this result comes from the fact that many authors restrict attention to the equatorial plane in order to simplify computations (e.g., \cite{Robledo11,Said10}). Note that the non-triviality of the results obtained in \cite{Robledo11,Said10} is the result of the non-zero accceleration of their chosen test particles.

Finally, we showed how, as an immediate consequence of Proposition \ref{DiracEquator}, spin precession is trivial for Dirac particles in Schwarschild geometry (Corollary \ref{DiracSchwarzschild}). 

\begin{figure}[p]
\caption{A counter-rotating orbit with $E=2,\Phi=3, \kappa=12$ and initial data $r(0)=20, \vartheta(0)=1.57$, and  $\phi(0)=0$.}
\begin{subfigure}[b]{0.99\textwidth}
\centering
\includegraphics[width=2in]{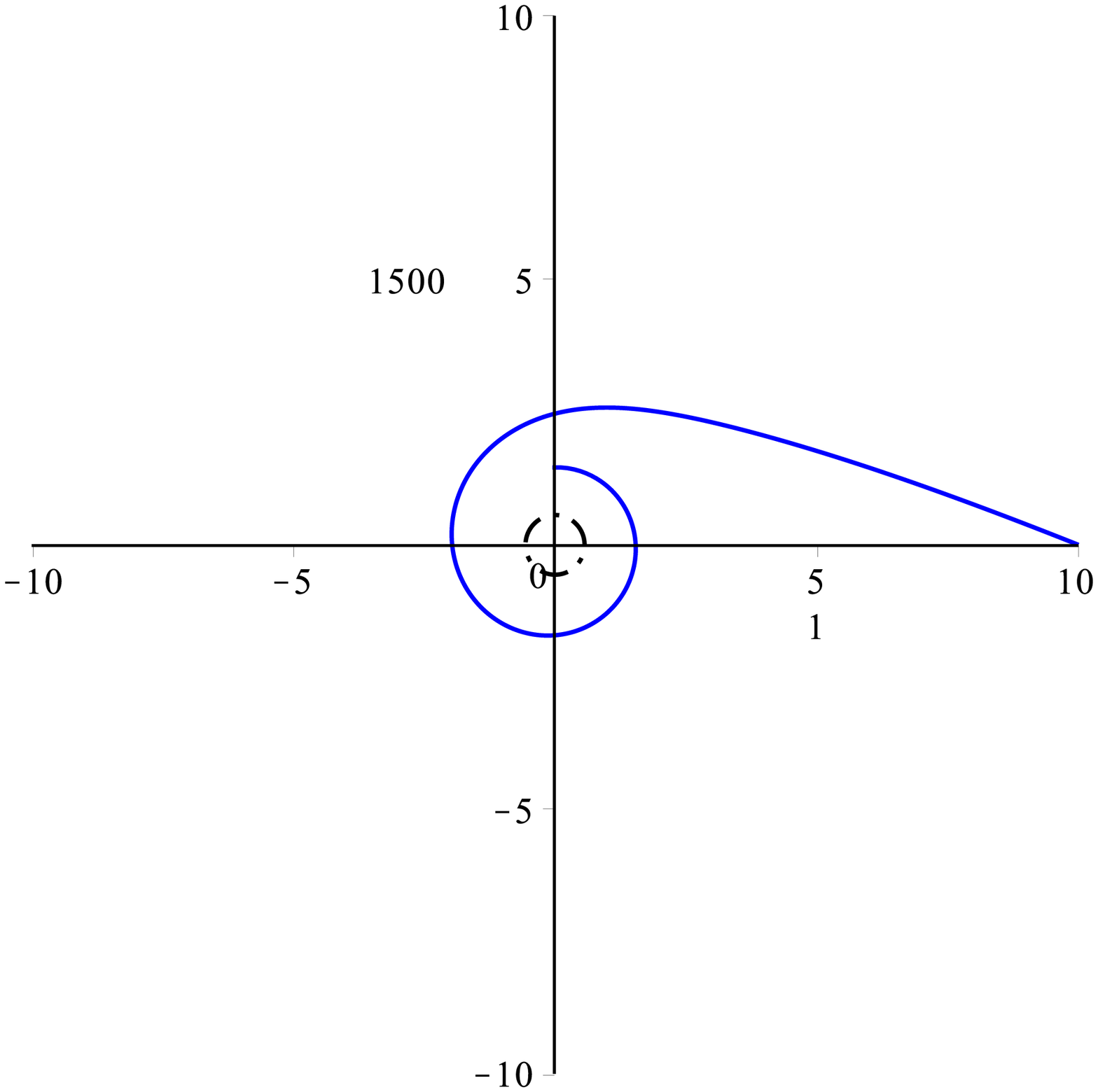}
\caption{The orbit in polar coordinates $(x=r\cos\phi,y=r\sin\phi).$}
\end{subfigure}
\\
\begin{subfigure}[b]{0.99\textwidth}
\centering
\includegraphics[width=2in]{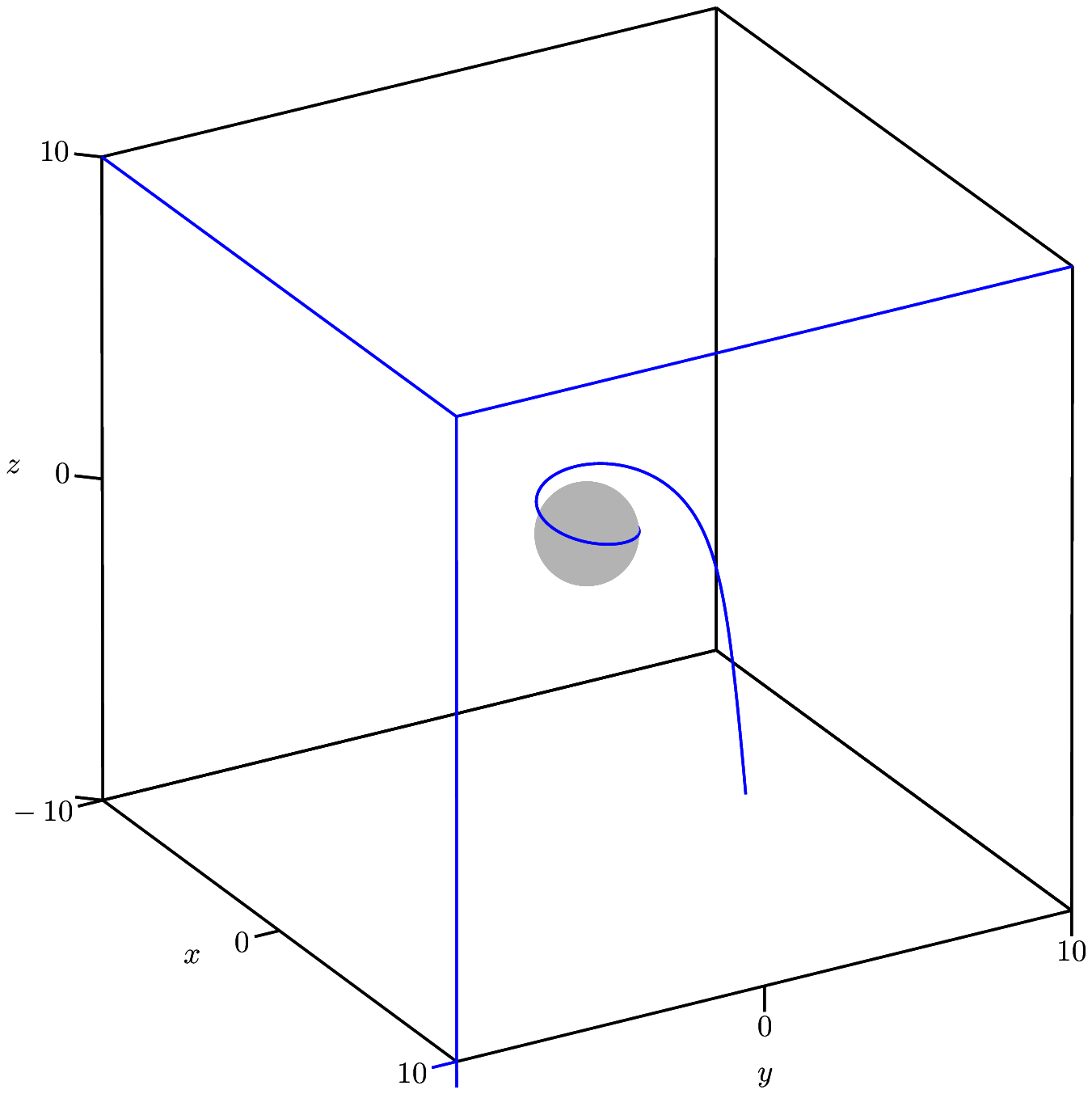}
\caption{The orbit in 3D spherical coordinates $\left(x=r\cos\phi\sin\vartheta, y=r\sin\phi\sin\vartheta, z=r\cos\vartheta\right)$. }
\end{subfigure}
\\
\begin{subfigure}[b]{0.99\textwidth}
\centering
\includegraphics[width=2in]{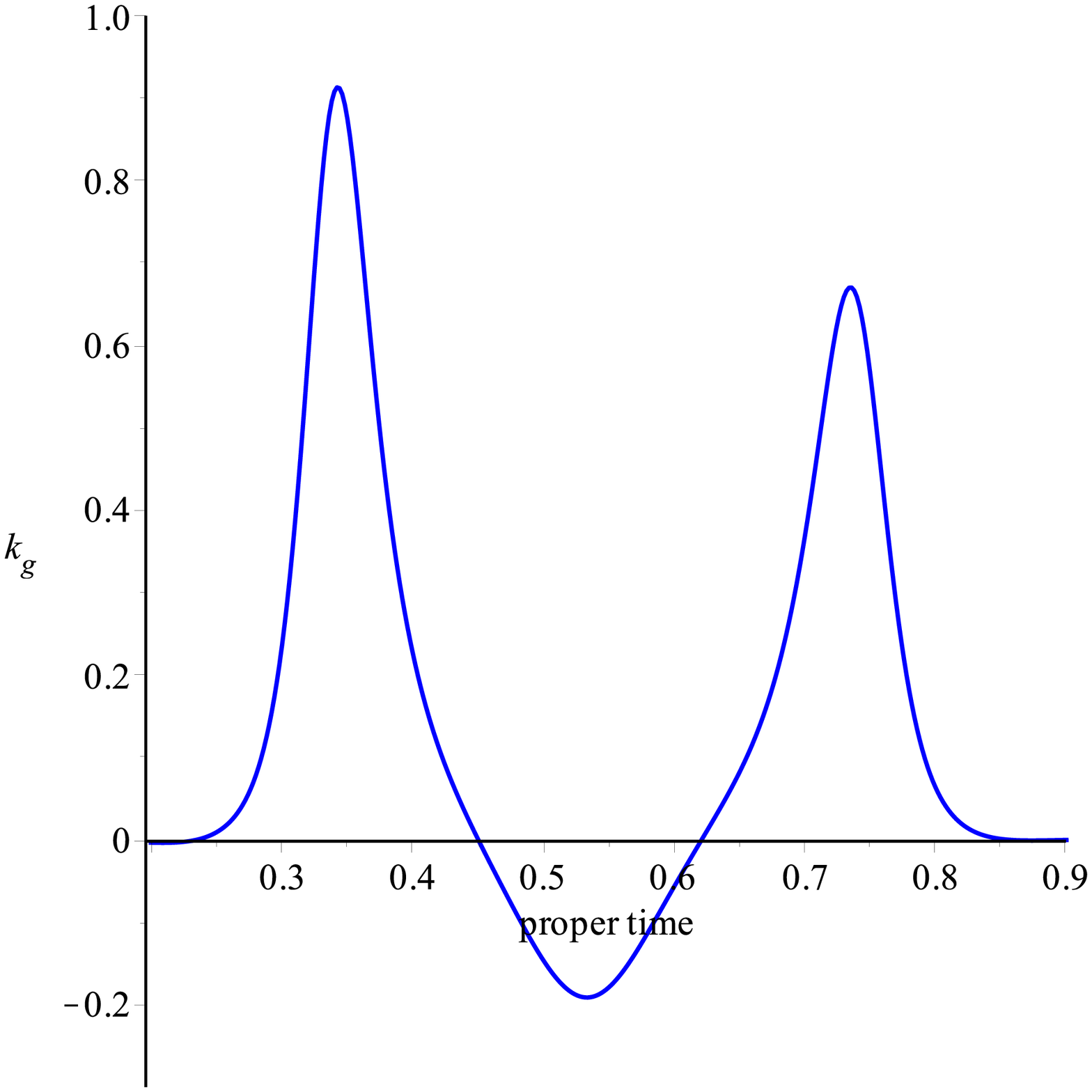}
\caption{The spherical curvature $k_{g}$ as a function of propertime.}
\end{subfigure}
\end{figure}
 
\begin{figure}[p]
\caption{A counter-rotating orbit with $E=1.004,\Phi=-4, \kappa=60$ and initial data $r(0)=20, \vartheta(0)=1.57$, and  $\phi(0)=0$.}
\begin{subfigure}[b]{0.99\textwidth}
\centering
\includegraphics[width=2in]{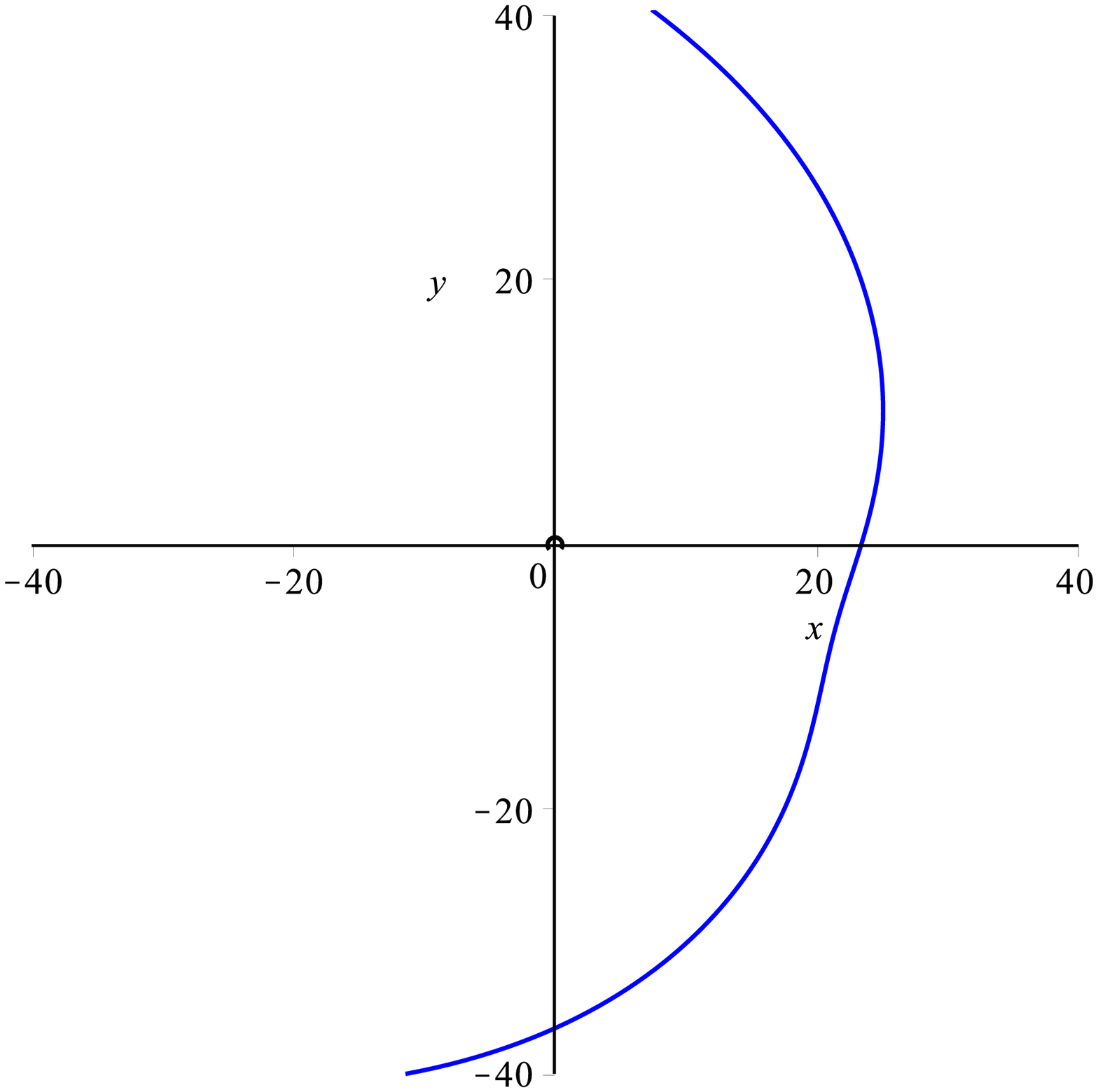}
\caption{The orbit in polar coordinates $(x=r\cos\phi,y=r\sin\phi).$}
\end{subfigure}
\\
\begin{subfigure}[b]{0.99\textwidth}
\centering
\includegraphics[width=2in]{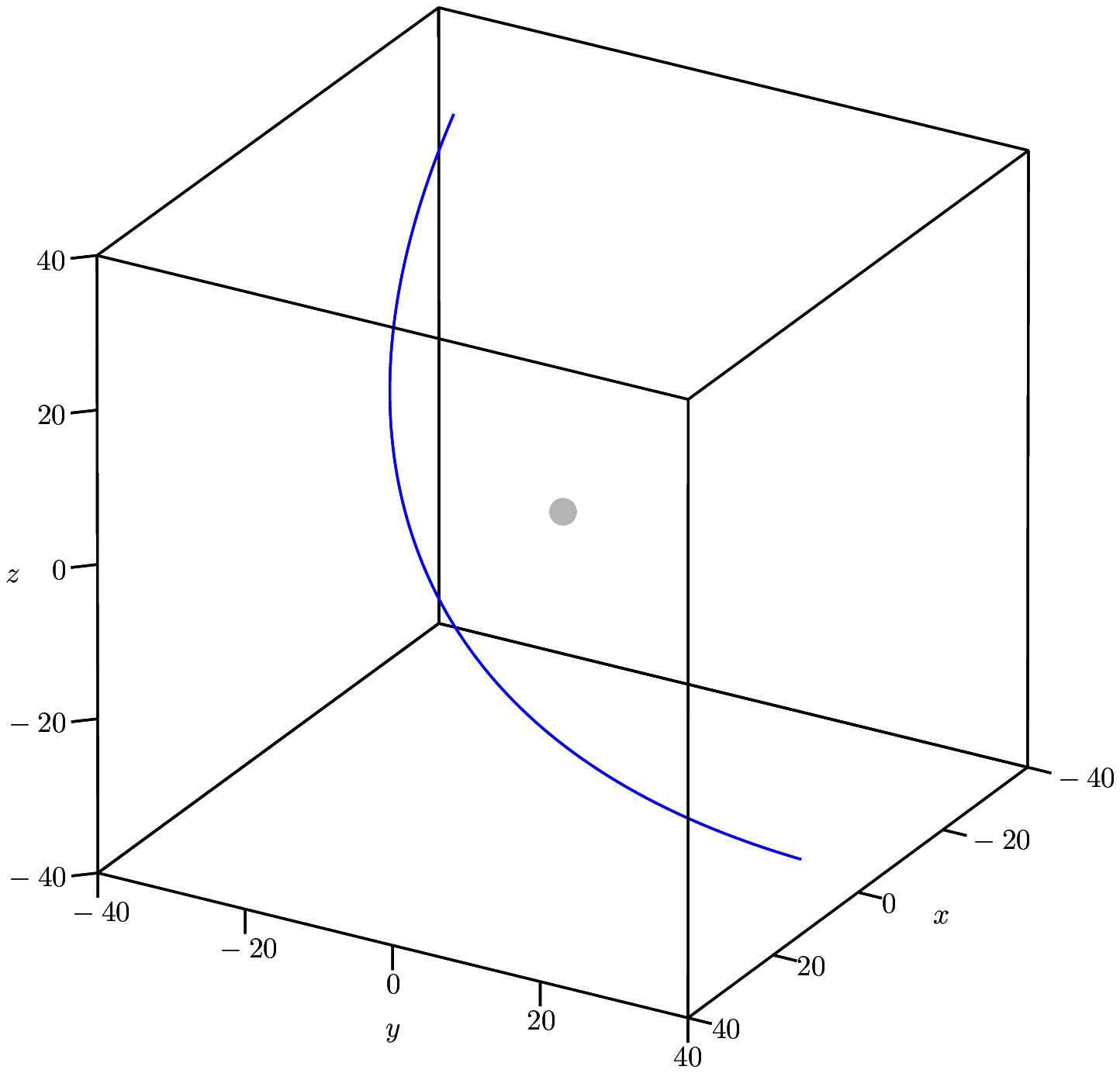}
\caption{The orbit in 3D spherical coordinates $\left(x=r\cos\phi\sin\vartheta, y=r\sin\phi\sin\vartheta, z=r\cos\vartheta\right)$. }
\end{subfigure}
\\
\begin{subfigure}[b]{0.99\textwidth}
\centering
\includegraphics[width=2in]{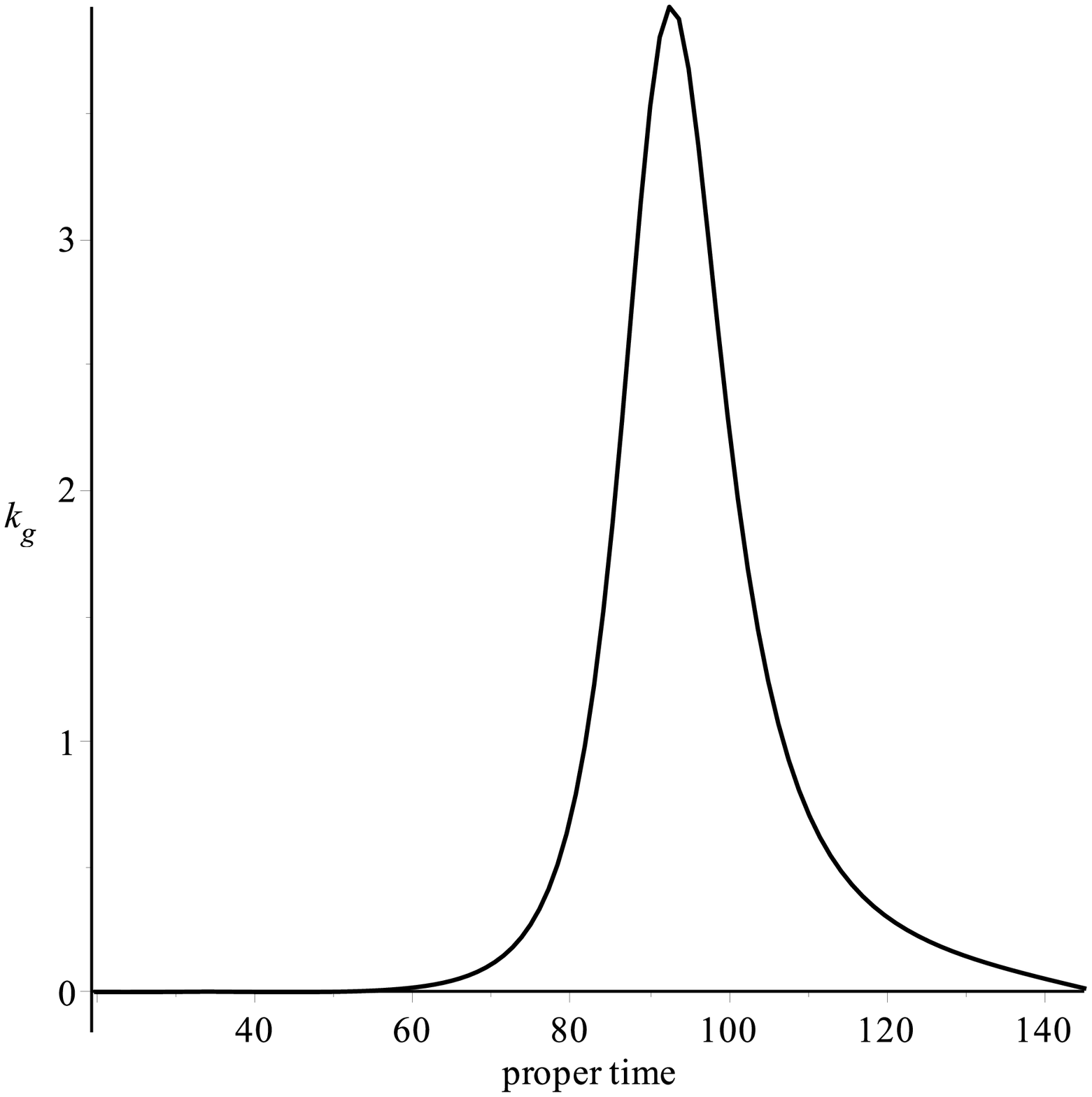}
\caption{The spherical curvature $k_{g}$ as a function of propertime.}
\end{subfigure}
\end{figure}

\begin{figure}[p]
\caption{A counter-rotating orbit with $E=1.004,\Phi=4, \kappa=16$ and initial data $r(0)=20, \vartheta(0)=1.57$, and  $\phi(0)=0$.}
\begin{subfigure}[b]{0.99\textwidth}
\centering
\includegraphics[width=2in]{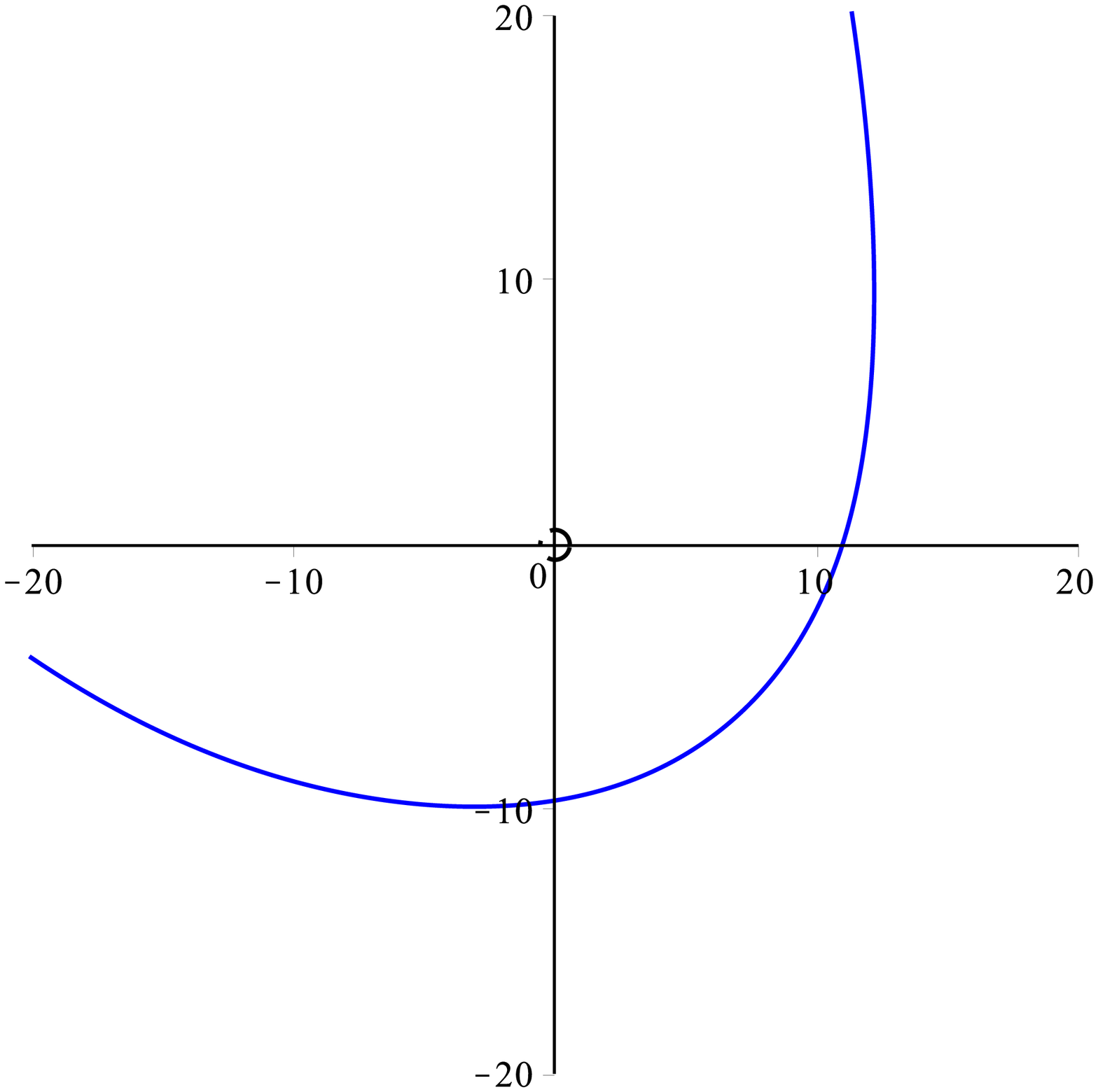}
\caption{The orbit in polar coordinates $(x=r\cos\phi,y=r\sin\phi).$}
\end{subfigure}
\\
\begin{subfigure}[b]{0.99\textwidth}
\centering
\includegraphics[width=2in]{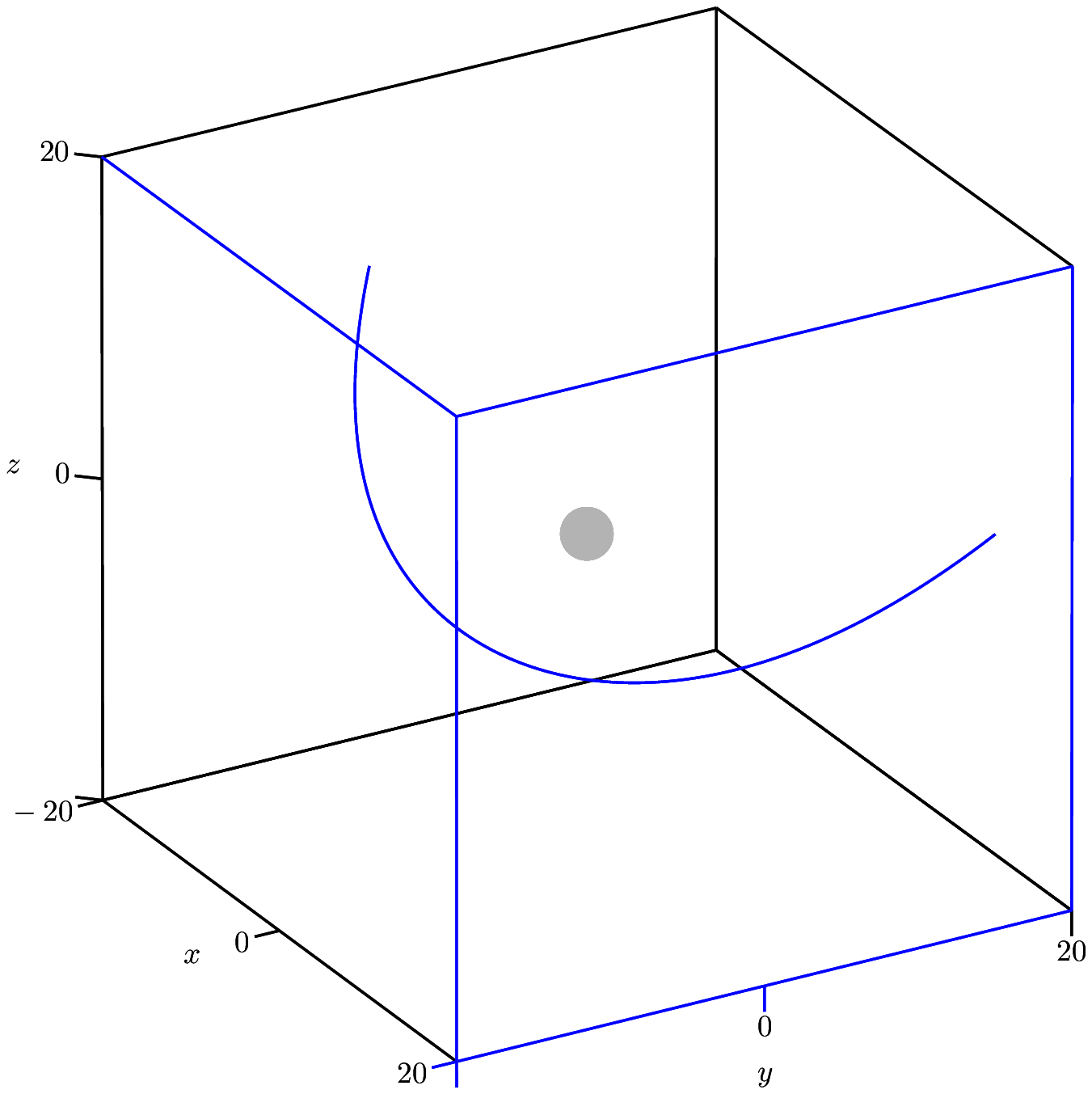}
\caption{The orbit in 3D spherical coordinates $\left(x=r\cos\phi\sin\vartheta, y=r\sin\phi\sin\vartheta, z=r\cos\vartheta\right)$. }
\end{subfigure}
\\
\begin{subfigure}[b]{0.99\textwidth}
\centering
\includegraphics[width=2in]{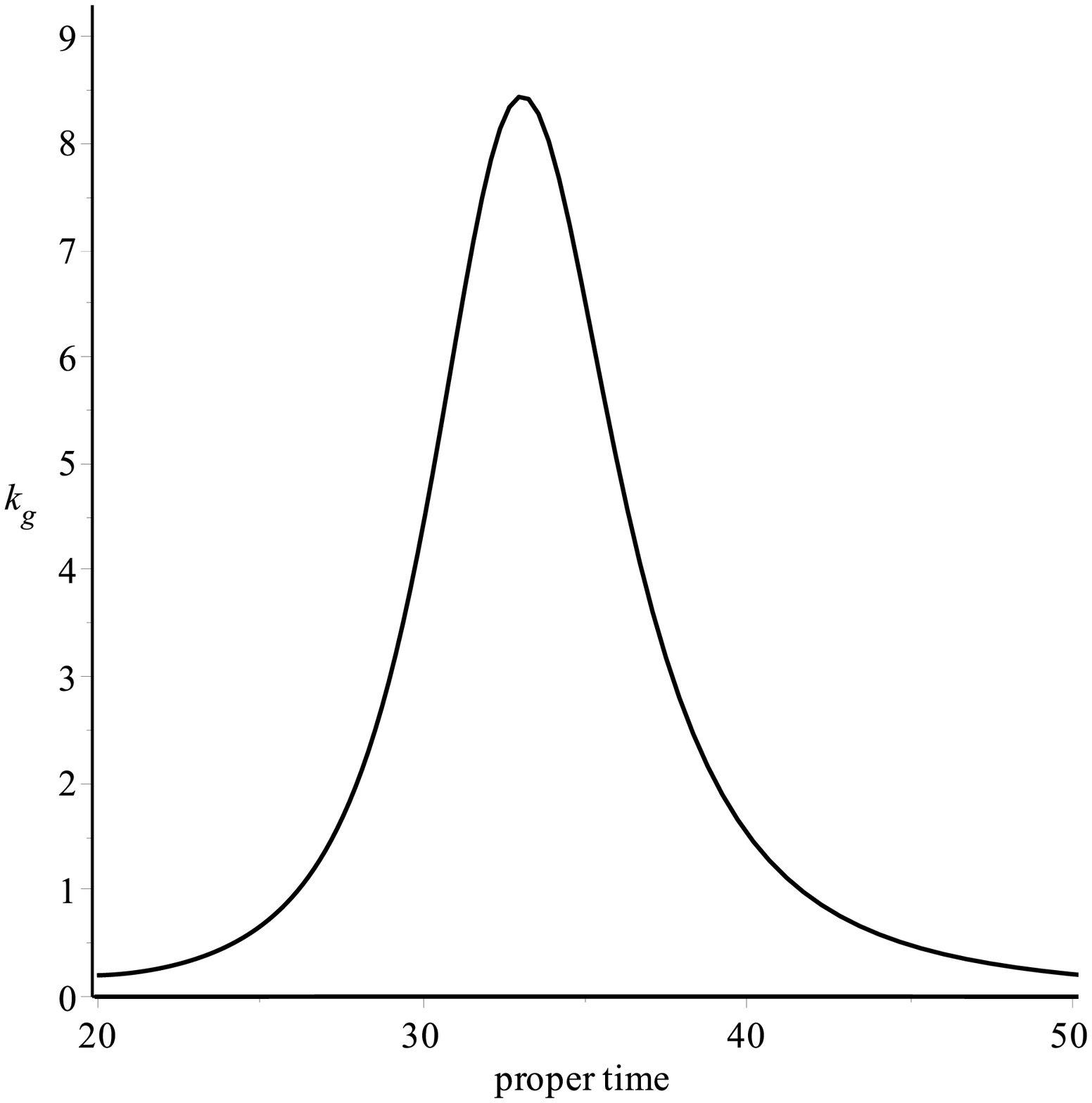}
\caption{The spherical curvature $k_{g}$ as a function of propertime.}
\end{subfigure}
\end{figure}

In order to qualitatively analyze the expression we have obtained for the spin precession of Dirac particles, we present some plots. In each of 3 sets of figures, the first figure (a) shows the
orbital behaviour of the timelike geodesic with $(r(s),\phi(s))$ as polar
coordinates, the second figure (b) depicts the same orbits in three
dimensions with spherical coordinates $(r(s),\vartheta(s),\phi(s))$, and the
last figure (c) depicts the spherical curvature $k_{g}$ as a function of propertime $\tau$. Figures 1 and 3 present co-rotating orbits, while Figure 2 presents a counter-rotating orbit. 
 
 \newpage
\bibliography{Bibilio}
\end{document}